\documentclass[preprint2]{aastex}
\slugcomment{Submitted to Astrophysical Journal Letters}
\def\fileversion{v1.20a}
\def\filedate{21.6.94}
\edef\epsfigRestoreAt{\catcode`@=\number\catcode`@\relax}%
\catcode`\@=11\relax
\ifx\undefined\@makeother                
\def\@makeother#1{\catcode`#1=12\relax}  
\fi                                      
\immediate\write16{Document style option `epsfig', \fileversion\space
<\filedate> (edited by SPQR + pks)}
\newcount\EPS@Height \newcount\EPS@Width \newcount\EPS@xscale
\newcount\EPS@yscale
\def\psfigdriver#1{%
  \bgroup\edef\next{\def\noexpand\tempa{#1}}%
    \uppercase\expandafter{\next}%
    \def\LN{DVITOLN03}%
    \def\DVItoPS{DVITOPS}%
    \def\DVIPS{DVIPS}%
    \def\emTeX{EMTEX}%
    \def\OzTeX{OZTEX}%
    \def\Textures{TEXTURES}%
    \global\chardef\fig@driver=0
    \ifx\tempa\LN
        \global\chardef\fig@driver=0\fi
    \ifx\tempa\DVItoPS
        \global\chardef\fig@driver=1\fi
    \ifx\tempa\DVIPS
        \global\chardef\fig@driver=2\fi
    \ifx\tempa\emTeX
        \global\chardef\fig@driver=3\fi
    \ifx\tempa\OzTeX
        \global\chardef\fig@driver=4\fi
    \ifx\tempa\Textures
        \global\chardef\fig@driver=5\fi
  \egroup
\def\psfig@start{}%
\def\psfig@end{}%
\def\epsfig@gofer{}%
\ifcase\fig@driver
\typeout{WARNING! ****
 no specials for LN03 psfig}%
\or 
\def\psfig@start{}%
\def\psfig@end{\special{dvitops: import \@p@sfilefinal \space
\@p@swidth sp \space \@p@sheight sp \space fill}%
\if@clip \typeout{Clipping not supported}\fi
\if@angle \typeout{Rotating not supported}\fi
}%
\let\epsfig@gofer\psfig@end
\or 
\def\psfig@start{\special{ps::[begin]  \@p@swidth \space \@p@sheight \space%
        \@p@sbbllx \space \@p@sbblly \space%
        \@p@sbburx \space \@p@sbbury \space%
        startTexFig \space }%
        \if@clip
                \if@verbose
                        \typeout{(clipped to BB) }%
                \fi
                \special{ps:: doclip \space }%
        \fi
        \if@angle              
                \special {ps:: \@p@sangle \space rotate \space}
        \fi
        \special{ps: plotfile \@p@sfilefinal \space }%
        \special{ps::[end] endTexFig \space }%
}%
\def\psfig@end{}%
\def\epsfig@gofer{\if@clip
                        \if@verbose
                           \typeout{(clipped to BB)}%
                        \fi
                        \epsfclipon
                  \fi
                  \epsfsetgraph{\@p@sfilefinal}%
}%
\or 
\typeout{WARNING. You must have a .bb info file with the Bounding Box
  of the pcx file}%
\def\psfig@start{}%
\def\psfig@end{\typeout{pcx import of \@p@sfilefinal}%
\if@clip \typeout{Clipping not supported}\fi
\if@angle \typeout{Rotating not supported}\fi
\raisebox{\@p@srheight sp}{\special{em: graph \@p@sfilefinal}}}%
\def\epsfig@gofer{}%
\or 
\def\psfig@start{}%
\def\psfig@end{%
\EPS@Width\@p@swidth
\EPS@Height\@p@sheight
\divide\EPS@Width by 65781  
\divide\EPS@Height by 65781
\special{epsf=\@p@sfilefinal
\space
width=\the\EPS@Width
\space
height=\the\EPS@Height
}%
\if@clip \typeout{Clipping not supported}\fi
\if@angle \typeout{Rotating not supported}\fi
}%
\let\epsfig@gofer\psfig@end
\or 
\def\psfig@end{
         \EPS@Width=\@bbw  
         \divide\EPS@Width by 1000
         \EPS@xscale=\@p@swidth \divide \EPS@xscale by \EPS@Width
         \EPS@Height=\@bbh  
         \divide\EPS@Height by 1000
         \EPS@yscale=\@p@sheight \divide \EPS@yscale by\EPS@Height
  \ifnum\EPS@xscale>\EPS@yscale\EPS@xscale=\EPS@yscale\fi
\if@clip
   \if@verbose
      \typeout{(clipped to BB)}%
   \fi
   \epsfclipon
\fi
\special{illustration \@p@sfilefinal\space scaled \the\EPS@xscale}%
}%
\def\psfig@start{}%
\let\epsfig\psfig
\else
\typeout{WARNING. *** unknown  driver - no psfig}%
\fi
}%
\newdimen\ps@dimcent
\ifx\undefined\fbox
\newdimen\fboxrule
\newdimen\fboxsep
\newdimen\ps@tempdima
\newbox\ps@tempboxa
\long\def\fbox#1{\leavevmode\setbox\ps@tempboxa\hbox{#1}\ps@tempdima\fboxrule
    \advance\ps@tempdima \fboxsep \advance\ps@tempdima \dp\ps@tempboxa
   \hbox{\lower \ps@tempdima\hbox
  {\vbox{\hrule height \fboxrule
          \hbox{\vrule width \fboxrule \hskip\fboxsep
          \vbox{\vskip\fboxsep \box\ps@tempboxa\vskip\fboxsep}\hskip
                 \fboxsep\vrule width \fboxrule}%
                 \hrule height \fboxrule}}}}%
\fi
\ifx\@ifundefined\undefined
\long\def\@ifundefined#1#2#3{\expandafter\ifx\csname
  #1\endcsname\relax#2\else#3\fi}%
\fi
\@ifundefined{typeout}%
{\gdef\typeout#1{\immediate\write\sixt@@n{#1}}}%
{\relax}%
\@ifundefined{epsfig}{}{\typeout{EPSFIG --- already loaded} }%
\@ifundefined{epsfbox}{\input epsf}{}%
\ifx\undefined\@latexerr
        \newlinechar`\^^J
        \def\@spaces{\space\space\space\space}%
        \def\@latexerr#1#2{%
        \edef\@tempc{#2}\expandafter\errhelp\expandafter{\@tempc}%
        \typeout{Error. \space see a manual for explanation.^^J
         \space\@spaces\@spaces\@spaces Type \space H <return> \space for
         immediate help.}\errmessage{#1}}%
\fi
\def\@whattodo{You tried to include a PostScript figure which
cannot be found^^JIf you press return to carry on anyway,^^J
The failed name will be printed in place of the figure.^^J
or type X to quit}%
\def\@whattodobb{You tried to include a PostScript figure which
has no^^Jbounding box, and you supplied none.^^J
If you press return to carry on anyway,^^J
The failed name will be printed in place of the figure.^^J
or type X to quit}%
\def\@nnil{\@nil}%
\def\@empty{}%
\def\@psdonoop#1\@@#2#3{}%
\def\@psdo#1:=#2\do#3{\edef\@psdotmp{#2}\ifx\@psdotmp\@empty \else
    \expandafter\@psdoloop#2,\@nil,\@nil\@@#1{#3}\fi}%
\def\@psdoloop#1,#2,#3\@@#4#5{\def#4{#1}\ifx #4\@nnil \else
       #5\def#4{#2}\ifx #4\@nnil \else#5\@ipsdoloop #3\@@#4{#5}\fi\fi}%
\def\@ipsdoloop#1,#2\@@#3#4{\def#3{#1}\ifx #3\@nnil
       \let\@nextwhile=\@psdonoop \else
      #4\relax\let\@nextwhile=\@ipsdoloop\fi\@nextwhile#2\@@#3{#4}}%
\def\@tpsdo#1:=#2\do#3{\xdef\@psdotmp{#2}\ifx\@psdotmp\@empty \else
    \@tpsdoloop#2\@nil\@nil\@@#1{#3}\fi}%
\def\@tpsdoloop#1#2\@@#3#4{\def#3{#1}\ifx #3\@nnil
       \let\@nextwhile=\@psdonoop \else
      #4\relax\let\@nextwhile=\@tpsdoloop\fi\@nextwhile#2\@@#3{#4}}%
\long\def\epsfaux#1#2:#3\\{\ifx#1\epsfpercent
   \def\testit{#2}\ifx\testit\epsfbblit
        \@atendfalse
        \epsf@atend #3 . \\%
        \if@atend
           \if@verbose
                \typeout{epsfig: found `(atend)'; continuing search}%
           \fi
        \else
                \epsfgrab #3 . . . \\%
                \epsffileokfalse\global\no@bbfalse
                \global\epsfbbfoundtrue
        \fi
   \fi\fi}%
\def\epsf@atendlit{(atend)}
\def\epsf@atend #1 #2 #3\\{%
   \def\epsf@tmp{#1}\ifx\epsf@tmp\empty
      \epsf@atend #2 #3 .\\\else
   \ifx\epsf@tmp\epsf@atendlit\@atendtrue\fi\fi}%

\chardef\trig@letter = 11
\chardef\other = 12
 
\newif\ifdebug 
\newif\ifc@mpute 
\newif\if@atend
\c@mputetrue 
 
\let\then = \relax
\def\r@dian{pt }%
\let\r@dians = \r@dian
\let\dimensionless@nit = \r@dian
\let\dimensionless@nits = \dimensionless@nit
\def\internal@nit{sp }%
\let\internal@nits = \internal@nit
\newif\ifstillc@nverging
\def \Mess@ge #1{\ifdebug \then \message {#1} \fi}%
 
{ 
        \catcode `\@ = \trig@letter
        \gdef \nodimen {\expandafter \n@dimen \the \dimen}%
        \gdef \term #1 #2 #3%
               {\edef \t@ {\the #1}
                \edef \t@@ {\expandafter \n@dimen \the #2\r@dian}%
                \t@rm {\t@} {\t@@} {#3}%
               }%
        \gdef \t@rm #1 #2 #3%
               {{%
                \count 0 = 0
                \dimen 0 = 1 \dimensionless@nit
                \dimen 2 = #2\relax
                \Mess@ge {Calculating term #1 of \nodimen 2}%
                \loop
                \ifnum  \count 0 < #1
                \then   \advance \count 0 by 1
                        \Mess@ge {Iteration \the \count 0 \space}%
                        \Multiply \dimen 0 by {\dimen 2}%
                        \Mess@ge {After multiplication, term = \nodimen 0}%
                        \Divide \dimen 0 by {\count 0}%
                        \Mess@ge {After division, term = \nodimen 0}%
                \repeat
                \Mess@ge {Final value for term #1 of
                                \nodimen 2 \space is \nodimen 0}%
                \xdef \Term {#3 = \nodimen 0 \r@dians}%
                \aftergroup \Term
               }}%
        \catcode `\p = \other
        \catcode `\t = \other
        \gdef \n@dimen #1pt{#1} 
}%
 
\def \Divide #1by #2{\divide #1 by #2} 
 
\def \Multiply #1by #2
       {{
        \count 0 = #1\relax
        \count 2 = #2\relax
        \count 4 = 65536
        \Mess@ge {Before scaling, count 0 = \the \count 0 \space and
                        count 2 = \the \count 2}%
        \ifnum  \count 0 > 32767 
        \then   \divide \count 0 by 4
                \divide \count 4 by 4
        \else   \ifnum  \count 0 < -32767
                \then   \divide \count 0 by 4
                        \divide \count 4 by 4
                \else
                \fi
        \fi
        \ifnum  \count 2 > 32767 
        \then   \divide \count 2 by 4
                \divide \count 4 by 4
        \else   \ifnum  \count 2 < -32767
                \then   \divide \count 2 by 4
                        \divide \count 4 by 4
                \else
                \fi
        \fi
        \multiply \count 0 by \count 2
        \divide \count 0 by \count 4
        \xdef \product {#1 = \the \count 0 \internal@nits}%
        \aftergroup \product
       }}%
 
\def\r@duce{\ifdim\dimen0 > 90\r@dian \then   
                \multiply\dimen0 by -1
                \advance\dimen0 by 180\r@dian
                \r@duce
            \else \ifdim\dimen0 < -90\r@dian \then  
                \advance\dimen0 by 360\r@dian
                \r@duce
                \fi
            \fi}%
 
\def\Sine#1%
       {{%
        \dimen 0 = #1 \r@dian
        \r@duce
        \ifdim\dimen0 = -90\r@dian \then
           \dimen4 = -1\r@dian
           \c@mputefalse
        \fi
        \ifdim\dimen0 = 90\r@dian \then
           \dimen4 = 1\r@dian
           \c@mputefalse
        \fi
        \ifdim\dimen0 = 0\r@dian \then
           \dimen4 = 0\r@dian
           \c@mputefalse
        \fi
        \ifc@mpute \then
                \divide\dimen0 by 180
                \dimen0=3.141592654\dimen0
                \dimen 2 = 3.1415926535897963\r@dian 
                \divide\dimen 2 by 2 
                \Mess@ge {Sin: calculating Sin of \nodimen 0}%
                \count 0 = 1 
                \dimen 2 = 1 \r@dian 
                \dimen 4 = 0 \r@dian 
                \loop
                        \ifnum  \dimen 2 = 0 
                        \then   \stillc@nvergingfalse
                        \else   \stillc@nvergingtrue
                        \fi
                        \ifstillc@nverging 
                        \then   \term {\count 0} {\dimen 0} {\dimen 2}%
                                \advance \count 0 by 2
                                \count 2 = \count 0
                                \divide \count 2 by 2
                                \ifodd  \count 2 
                                \then   \advance \dimen 4 by \dimen 2
                                \else   \advance \dimen 4 by -\dimen 2
                                \fi
                \repeat
        \fi
                        \xdef \sine {\nodimen 4}%
       }}%
 
\def\Cosine#1{\ifx\sine\UnDefined\edef\Savesine{\relax}\else
                             \edef\Savesine{\sine}\fi
        {\dimen0=#1\r@dian\multiply\dimen0 by -1
         \advance\dimen0 by 90\r@dian
         \Sine{\nodimen 0}%
         \xdef\cosine{\sine}%
         \xdef\sine{\Savesine}}}
\def\psdraft{\def\@psdraft{0}}%
\def\psfull{\def\@psdraft{1}}%
\psfull
\newif\if@compress
\def\pscompress{\@compresstrue}
\def\psnocompress{\@compressfalse}
\@compressfalse
\newif\if@scalefirst
\def\psscalefirst{\@scalefirsttrue}%
\def\psrotatefirst{\@scalefirstfalse}%
\psrotatefirst
\newif\if@draftbox
\def\psnodraftbox{\@draftboxfalse}%
\@draftboxtrue
\newif\if@noisy
\@noisyfalse
\newif\ifno@bb
\newif\if@bbllx
\newif\if@bblly
\newif\if@bburx
\newif\if@bbury
\newif\if@height
\newif\if@width
\newif\if@rheight
\newif\if@rwidth
\newif\if@angle
\newif\if@clip
\newif\if@verbose
\newif\if@prologfile
\def\@p@@sprolog#1{\@prologfiletrue\def\@prologfileval{#1}}%
\def\@p@@sclip#1{\@cliptrue}%
\newif\ifepsfig@dos  
\def\epsfigdos{\epsfig@dostrue}%
\epsfig@dosfalse
\newif\ifuse@psfig
\def\ParseName#1{\expandafter\@Parse#1}%
\def\@Parse#1.#2:{\gdef\BaseName{#1}\gdef\FileType{#2}}%

\def\@p@@sfile#1{%
  \ifepsfig@dos
     \ParseName{#1:}%
  \else
     \gdef\BaseName{#1}\gdef\FileType{}%
  \fi
  \def\@p@sfile{NO FILE: #1}%
  \def\@p@sfilefinal{NO FILE: #1}%
  \openin1=#1
  \ifeof1\closein1\openin1=\BaseName.bb
    \ifeof1\closein1
      \if@bbllx                 
        \if@bblly\if@bburx\if@bbury
          \def\@p@sfile{#1}%
          \def\@p@sfilefinal{#1}%
        \fi\fi\fi
      \else                     
        \@latexerr{ERROR. PostScript file #1 not found}\@whattodo
        \@p@@sbbllx{100bp}%
        \@p@@sbblly{100bp}%
        \@p@@sbburx{200bp}%
        \@p@@sbbury{200bp}%
        \psdraft
      \fi
    \else                       
      \closein1%
      \edef\@p@sfile{\BaseName.bb}%
      \typeout{using BB from \@p@sfile}%
      \ifnum\fig@driver=3
        \edef\@p@sfilefinal{\BaseName.pcx}%
      \else
        \ifepsfig@dos
          \edef\@p@sfilefinal{"`gunzip -c `texfind \BaseName.{z,Z,gz}"}%
        \else
          \edef\@p@sfilefinal{"`epsfig \if@compress-c \fi#1"}%
        \fi
      \fi
    \fi
  \else\closein1                
    \edef\@p@sfile{#1}%
    \if@compress  
      \edef\@p@sfilefinal{"`epsfig -c #1"}%
    \else
      \edef\@p@sfilefinal{#1}%
    \fi
  \fi%
}

\let\@p@@sfigure\@p@@sfile
\def\@p@@sbbllx#1{%
                                            \@bbllxtrue
                \ps@dimcent=#1
                \edef\@p@sbbllx{\number\ps@dimcent}%
                \divide\ps@dimcent by65536
                \global\edef\epsfllx{\number\ps@dimcent}%
}%
\def\@p@@sbblly#1{%
                \@bbllytrue
                \ps@dimcent=#1
                \edef\@p@sbblly{\number\ps@dimcent}%
                \divide\ps@dimcent by65536
                \global\edef\epsflly{\number\ps@dimcent}%
}%
\def\@p@@sbburx#1{%
                \@bburxtrue
                \ps@dimcent=#1
                \edef\@p@sbburx{\number\ps@dimcent}%
                \divide\ps@dimcent by65536
                \global\edef\epsfurx{\number\ps@dimcent}%
}%
\def\@p@@sbbury#1{%
                \@bburytrue
                \ps@dimcent=#1
                \edef\@p@sbbury{\number\ps@dimcent}%
                \divide\ps@dimcent by65536
                \global\edef\epsfury{\number\ps@dimcent}%
}%
\def\@p@@sheight#1{%
                \@heighttrue
                \global\epsfysize=#1
                \ps@dimcent=#1
                \edef\@p@sheight{\number\ps@dimcent}%
}%
\def\@p@@swidth#1{%
                \@widthtrue
                \global\epsfxsize=#1
                \ps@dimcent=#1
                \edef\@p@swidth{\number\ps@dimcent}%
}%
\def\@p@@srheight#1{%
                \@rheighttrue\use@psfigtrue
                \ps@dimcent=#1
                \edef\@p@srheight{\number\ps@dimcent}%
}%
\def\@p@@srwidth#1{%
                \@rwidthtrue\use@psfigtrue
                \ps@dimcent=#1
                \edef\@p@srwidth{\number\ps@dimcent}%
}%
\def\@p@@sangle#1{%
                \use@psfigtrue
                \@angletrue
                \edef\@p@sangle{#1}%
}%
\def\@p@@ssilent#1{%
                \@verbosefalse
}%
\def\@p@@snoisy#1{%
                \@verbosetrue
}%
\def\@cs@name#1{\csname #1\endcsname}%
\def\@setparms#1=#2,{\@cs@name{@p@@s#1}{#2}}%
\def\ps@init@parms{%
                \@bbllxfalse \@bbllyfalse
                \@bburxfalse \@bburyfalse
                \@heightfalse \@widthfalse
                \@rheightfalse \@rwidthfalse
                \def\@p@sbbllx{}\def\@p@sbblly{}%
                \def\@p@sbburx{}\def\@p@sbbury{}%
                \def\@p@sheight{}\def\@p@swidth{}%
                \def\@p@srheight{}\def\@p@srwidth{}%
                \def\@p@sangle{0}%
                \def\@p@sfile{}%
                \use@psfigfalse
                \@prologfilefalse
                \def\@sc{}%
                \if@noisy
                        \@verbosetrue
                \else
                        \@verbosefalse
                \fi
                \@clipfalse
}%
\def\parse@ps@parms#1{%
                \@psdo\@psfiga:=#1\do
                   {\expandafter\@setparms\@psfiga,}%
\if@prologfile
\fi
}%
\def\bb@missing{%
        \if@verbose
            \typeout{psfig: searching \@p@sfile \space  for bounding box}%
        \fi
        \epsfgetbb{\@p@sfile}%
        \ifepsfbbfound
            \ps@dimcent=\epsfllx bp\edef\@p@sbbllx{\number\ps@dimcent}%
            \ps@dimcent=\epsflly bp\edef\@p@sbblly{\number\ps@dimcent}%
            \ps@dimcent=\epsfurx bp\edef\@p@sbburx{\number\ps@dimcent}%
            \ps@dimcent=\epsfury bp\edef\@p@sbbury{\number\ps@dimcent}%
        \else
            \epsfbbfoundfalse
        \fi
}
\newdimen\p@intvaluex
\newdimen\p@intvaluey
\def\rotate@#1#2{{\dimen0=#1 sp\dimen1=#2 sp
                  \global\p@intvaluex=\cosine\dimen0
                  \dimen3=\sine\dimen1
                  \global\advance\p@intvaluex by -\dimen3
                  \global\p@intvaluey=\sine\dimen0
                  \dimen3=\cosine\dimen1
                  \global\advance\p@intvaluey by \dimen3
                  }}%
\def\compute@bb{%
                \epsfbbfoundfalse
                \if@bbllx\epsfbbfoundtrue\fi
                \if@bblly\epsfbbfoundtrue\fi
                \if@bburx\epsfbbfoundtrue\fi
                \if@bbury\epsfbbfoundtrue\fi
                \ifepsfbbfound\else\bb@missing\fi
                \ifepsfbbfound\else
                \@latexerr{ERROR. cannot locate BoundingBox}\@whattodobb
                        \@p@@sbbllx{100bp}%
                        \@p@@sbblly{100bp}%
                        \@p@@sbburx{200bp}%
                        \@p@@sbbury{200bp}%
                        \no@bbtrue
                        \psdraft
                \fi
                \count203=\@p@sbburx
                \count204=\@p@sbbury
                \advance\count203 by -\@p@sbbllx
                \advance\count204 by -\@p@sbblly
                \edef\ps@bbw{\number\count203}%
                \edef\ps@bbh{\number\count204}%
                 \edef\@bbw{\number\count203}%
                \edef\@bbh{\number\count204}%
               \if@angle
                        \Sine{\@p@sangle}\Cosine{\@p@sangle}%
 
{\ps@dimcent=\maxdimen\xdef\r@p@sbbllx{\number\ps@dimcent}%
 
\xdef\r@p@sbblly{\number\ps@dimcent}%
 
\xdef\r@p@sbburx{-\number\ps@dimcent}%
 
\xdef\r@p@sbbury{-\number\ps@dimcent}}%
                        \def\minmaxtest{%
                           \ifnum\number\p@intvaluex<\r@p@sbbllx
                              \xdef\r@p@sbbllx{\number\p@intvaluex}\fi
                           \ifnum\number\p@intvaluex>\r@p@sbburx
                              \xdef\r@p@sbburx{\number\p@intvaluex}\fi
                           \ifnum\number\p@intvaluey<\r@p@sbblly
                              \xdef\r@p@sbblly{\number\p@intvaluey}\fi
                           \ifnum\number\p@intvaluey>\r@p@sbbury
                              \xdef\r@p@sbbury{\number\p@intvaluey}\fi
                           }%
                        \rotate@{\@p@sbbllx}{\@p@sbblly}%
                        \minmaxtest
                        \rotate@{\@p@sbbllx}{\@p@sbbury}%
                        \minmaxtest
                        \rotate@{\@p@sbburx}{\@p@sbblly}%
                        \minmaxtest
                        \rotate@{\@p@sbburx}{\@p@sbbury}%
                        \minmaxtest
 
\edef\@p@sbbllx{\r@p@sbbllx}\edef\@p@sbblly{\r@p@sbblly}%
 
\edef\@p@sbburx{\r@p@sbburx}\edef\@p@sbbury{\r@p@sbbury}%
                \fi
                \count203=\@p@sbburx
                \count204=\@p@sbbury
                \advance\count203 by -\@p@sbbllx
                \advance\count204 by -\@p@sbblly
                \edef\@bbw{\number\count203}%
                \edef\@bbh{\number\count204}%
}%
\def\in@hundreds#1#2#3{\count240=#2 \count241=#3
                     \count100=\count240        
                     \divide\count100 by \count241
                     \count101=\count100
                     \multiply\count101 by \count241
                     \advance\count240 by -\count101
                     \multiply\count240 by 10
                     \count101=\count240        
                     \divide\count101 by \count241
                     \count102=\count101
                     \multiply\count102 by \count241
                     \advance\count240 by -\count102
                     \multiply\count240 by 10
                     \count102=\count240        
                     \divide\count102 by \count241
                     \count200=#1\count205=0
                     \count201=\count200
                        \multiply\count201 by \count100
                        \advance\count205 by \count201
                     \count201=\count200
                        \divide\count201 by 10
                        \multiply\count201 by \count101
                        \advance\count205 by \count201
                     \count201=\count200
                        \divide\count201 by 100
                        \multiply\count201 by \count102
                        \advance\count205 by \count201
                     \edef\@result{\number\count205}%
}%
\def\compute@wfromh{%
                \in@hundreds{\@p@sheight}{\@bbw}{\@bbh}%
                \edef\@p@swidth{\@result}%
}%
\def\compute@hfromw{%
                \in@hundreds{\@p@swidth}{\@bbh}{\@bbw}%
                \edef\@p@sheight{\@result}%
}%
\def\compute@handw{%
                \if@height
                        \if@width
                        \else
                                \compute@wfromh
                        \fi
                \else
                        \if@width
                                \compute@hfromw
                        \else
                                \edef\@p@sheight{\@bbh}%
                                \edef\@p@swidth{\@bbw}%
                        \fi
                \fi
}%
\def\compute@resv{%
                \if@rheight \else \edef\@p@srheight{\@p@sheight} \fi
                \if@rwidth \else \edef\@p@srwidth{\@p@swidth} \fi
}%
\def\compute@sizes{%
        \if@scalefirst\if@angle
        \if@width
           \in@hundreds{\@p@swidth}{\@bbw}{\ps@bbw}%
           \edef\@p@swidth{\@result}%
        \fi
        \if@height
           \in@hundreds{\@p@sheight}{\@bbh}{\ps@bbh}%
           \edef\@p@sheight{\@result}%
        \fi
        \fi\fi
        \compute@handw
        \compute@resv
}
\long\def\graphic@verb#1{\def\next{#1}%
  {\expandafter\graphic@strip\meaning\next}}
\def\graphic@strip#1>{}
\def\graphic@zapspace#1{%
  #1\ifx\graphic@zapspace#1\graphic@zapspace%
  \else\expandafter\graphic@zapspace%
  \fi}
\def\psfig#1{%
\edef\@tempa{\graphic@zapspace#1{}}%
\ifvmode\leavevmode\fi\vbox {%
        \ps@init@parms
        \parse@ps@parms{\@tempa}%
        \ifnum\@psdraft=1
                \typeout{[\@p@sfilefinal]}%
                \if@verbose
                        \typeout{epsfig: using PSFIG macros}%
                \fi
                \psfig@method
        \else
                \epsfig@draft
        \fi
}
}%
\def\graphic@zapspace#1{%
  #1\ifx\graphic@zapspace#1\graphic@zapspace%
  \else\expandafter\graphic@zapspace%
  \fi}
\def\epsfig#1{%
\edef\@tempa{\graphic@zapspace#1{}}%
\ifvmode\leavevmode\fi\vbox {%
        \ps@init@parms
        \parse@ps@parms{\@tempa}%
        \ifnum\@psdraft=1
          \if@angle\use@psfigtrue\fi
          {\ifnum\fig@driver=1\global\use@psfigtrue\fi}%
          {\ifnum\fig@driver=3\global\use@psfigtrue\fi}%
          {\ifnum\fig@driver=4\global\use@psfigtrue\fi}%
          {\ifnum\fig@driver=5\global\use@psfigtrue\fi}%
                \ifuse@psfig
                        \if@verbose
                                \typeout{epsfig: using PSFIG macros}%
                        \fi
                        \psfig@method
                \else
                        \if@verbose
                                \typeout{epsfig: using EPSF macros}%
                        \fi
                        \epsf@method
                \fi
        \else
                \epsfig@draft
        \fi
}%
}%

\def\epsf@method{%
        \epsfbbfoundfalse
        \if@bbllx\epsfbbfoundtrue\fi
        \if@bblly\epsfbbfoundtrue\fi
        \if@bburx\epsfbbfoundtrue\fi
        \if@bbury\epsfbbfoundtrue\fi
        \ifepsfbbfound\else\epsfgetbb{\@p@sfile}\fi
        \ifepsfbbfound
           \typeout{<\@p@sfilefinal>}%
           \epsfig@gofer
        \else
          \@latexerr{ERROR - Cannot locate BoundingBox}\@whattodobb
          \@p@@sbbllx{100bp}%
          \@p@@sbblly{100bp}%
          \@p@@sbburx{200bp}%
          \@p@@sbbury{200bp}%
                \count203=\@p@sbburx
                \count204=\@p@sbbury
                \advance\count203 by -\@p@sbbllx
                \advance\count204 by -\@p@sbblly
                \edef\@bbw{\number\count203}%
                \edef\@bbh{\number\count204}%
          \compute@sizes
          \epsfig@@draft
       \fi
}%
\def\psfig@method{%
        \compute@bb
        \ifepsfbbfound
          \compute@sizes
          \psfig@start
          \vbox to \@p@srheight sp{\hbox to \@p@srwidth 
            sp{\hss}\vss\psfig@end}%
        \else
           \epsfig@draft
        \fi
}%
\def\epsfig@draft{\compute@bb\compute@sizes\epsfig@@draft}%
\def\epsfig@@draft{%
\typeout{<(draft only) \@p@sfilefinal>}%
\if@draftbox
        \hbox{{\fboxsep0pt\fbox{\vbox to \@p@srheight sp{%
        \vss\hbox to \@p@srwidth sp{ \hss 
           \expandafter\Literally\@p@sfilefinal\@nil
                          \hss }\vss
        }}}}%
\else
        \vbox to \@p@srheight sp{%
        \vss\hbox to \@p@srwidth sp{\hss}\vss}%
\fi
}%
\def\Literally#1\@nil{{\tt\graphic@verb{#1}}}
\psfigdriver{dvips}%
\epsfigRestoreAt

\lefthead{ }
\righthead{ }

\def \m3{{\rm Mark III}}
\def \etal {{\it et al.\ }}

\newcommand{\zi}{ \mbox{\boldmath$\xi$} }
\newcommand{\Bl}{ \left( }
\newcommand{\Br}{ \right)}

\begin{document}

\title{Dark Halo Shapes and the Fate of Stellar Bars}

\author{Amr  El-Zant}
\affil{Center for Astrophysics \& Space Astronomy, Campus Box 391, University
   of Colorado, Boulder, CO 80309-0391 and\\ Department of Physics \&
    Astronomy, University of Kentucky, Lexington, KY 40506-0055, USA \\
    email: {\tt elzant@pa.uky.edu}}  

\and
\author{Isaac Shlosman\altaffilmark{1}}
\affil{Joint Institute for Laboratory Astrophysics, University of Colorado,
   Campus Box 440, Boulder, CO 80309-0440, USA \\ email: {\tt
   shlosman@jila.colorado.edu}}      

\altaffiltext{1}{JILA Visiting Fellow. Permanent address: Department of
Physics \& Astronomy, University of Kentucky, Lexington, KY 40506-0055}

\begin{abstract}
We investigate the stability properties of trajectories in barred galaxies
with mildly triaxial halos by means of Liapunov exponents. This method is
perfectly suitable for time-dependent 3-D potentials where surfaces of sections
and other simple diagnostics are not applicable. We find that when halos are
centrally-concentrated most trajectories starting near the plane containing
the bar become chaotic. The spatial density distribution of these orbits does not
match that of the bar, being overextended in- and out-of-the plane compared to
the latter.  Moreover, the shape of many of the remaining regular trajectories
do not match the the bar density distribution, being too round.  Therefore,
time-independent self-consistent solutions are highly unlikely to be found.
When the non-rotating non-axisymmetric perturbation in the  potential reaches
$10\%$, almost all trajectories integrated are chaotic and have large Liapunov
exponents. No regular trajectories aligned with the bar have been found.
Hence, if the evolution of the density figure is directly related to the
characteristic  timescale of orbital instability,  bar dissolution would take
place on a timescale of few dynamical times. The slowly rotating non-axisymmetric
contribution to the potential required for the {\em onset} of widespread
chaotic behavior is remarkably small. Even  a potential axis ratio of $0.99$
results in large connected chaotic  regions dominating the space of initial
conditions. Systems consisting of centrally-concentrated axisymmetric halos
and stellar bars thus appear to be structurally unstable, and small ($\sim
1\%$) deviations from perfect axisymmetry should result in a bar dissolution
on a timescale significantly smaller than the Hubble time. Since halos found
in cold dark matter simulations of large scale structure are both
centrally-concentrated and triaxial it is unlikely that stellar bars embedded
in such halos would form and survive unless the halos are modified during the 
formation of the baryonic component. 
\end{abstract}


\bigskip 

\section{Introduction and motivation}

Stellar bars provide a significant impetus for dynamical and secular
evolution of disk galaxies. The main reason for this is that the
breaking of  axial
symmetry introduces gravitational torques, whose action can be described in
terms of a nonlocal viscosity (e.g., Larson 1984; Shlosman 1991). 
This causes accelerated redistribution of mass and angular momentum. 
Disk  -- halo interaction is also increased dramatically 
in the presence of a bar. This is particularly so
if the halo is also triaxial with significant contribution to the density
in the inner regions --- that is if it is centrally concentrated.
Such a strong interaction
raises questions about stability of the least massive object in this
configuration, the stellar bar. It also has clear implications on disk
formation and morphological evolution from high redshifts down to the local
universe.   In this paper, we analyze the orbital stability of trajectories
evolving in the potential of  a stellar 
bar embedded in massive halos of
various central concentrations and asymmetries
and discuss a number of corollaries, in order
to infer the overall stability of the bar.

Triaxial systems  must be built by dynamical trajectories that conserve, at
least approximately, invariants of motion, if they are to remain in
quasi-steady state.  This is necessary if their configuration (i.e.,
physical) space density is to  match  the triaxial shape of the system.
For example, a nonrotating steady state system, whose distribution function
depends only on the energy, is necessarily  spherical. While this is the only
``global'' integral of motion that always exists in time-independent systems, 
it is not sufficient to maintain a triaxial shape. Moreover, additional
 global integrals of motion 
are often associated  with spatial symmetries, which are lacking in triaxial
systems, rendering the problem of finding self-consistent solutions highly 
non-trivial (for a discussion of these issues for the case of 
triaxial elliptical galaxies  see, e.g., Merritt \& Fridman 1996
and Merritt \& Valluri 1996).
In some cases, symmetries different from simple spatial
ones  can exist, leading to  global integrals of motions that can
maintain the triaxial structure. Thus is the situation  with Stackel
potentials (e.g., de Zeeuw 1985). In the core region of such potentials the
effective symmetry is the near-homogeneity of the density distribution. The
potential can then be approximated as a quadratic form where the motion is
separable in Cartesian coordinates (in fact this type of system is not only
separable but even linear). Generic centrally-concentrated potentials do not
have such symmetries, the oscillations in the different degrees of freedom are,
in general, coupled and no global integrals of motion exist.
 This makes plausible a situation whereas  most trajectories in the 
 central  regions conserve only energy, in which case  self-consistent
solutions become impossible.

Centrally-concentrated mass distributions have been invoked by several authors 
in proposing a mechanism for secular evolution in galaxies, both in the case
of slowly  rotating elliptical galaxies (e.g., Norman, May \& van Albada 1985;
Merritt \& Fridman 1996; Holley-Bokelmann et al. 2002) and, in the context of
rapidly rotating bars, in disk galaxies (e.g., Pfenniger \& Norman 1990;
Norman,  Sellwood \& Hasan 1996).  For example, it has been argued that barred
galaxies that are not initially centrally-concentrated may acquire a central
mass concentration by accreting gas into the central region. This
creates the necessary coupling between the degrees of freedom so as to
destroy the integrals of motion for a large enough fraction of orbits and
dissolve the bar --- which is then replaced by a bulge-like structure. It
appears, however, that the large central masses required for such a scenario to
work, specifically the central black holes, are not confirmed by observations
(in preparation). An analogous situation should however transpire if the
coupling between  the degrees of freedom is mediated by the existence of  a
centrally-concentrated halo dominating the  inner density distribution. This
would be  the case for  systems with halos of the type found in cosmological
simulations of the cold dark matter (CDM) scenario of structure formation
(e.g., Navarro, Frenk \& White 1997; Moore \etal 1999).

Halos identified in cosmological  simulations are also invariably
found to be   triaxial (e.g., Warren et al 1992; Cole \& Lacey 1996). 
In general, the halo being built of non-dissipative material, will be much
more slowly rotating than an embedded baryonic bar formed in its central
region. The introduction of a bar, i.e., an additional triaxial
configuration, thus ensures that even energy is not conserved in any uniformly
rotating  frame of reference. The loss of an additional global (time
translation) symmetry makes it even  more likely  that a majority of
trajectories will be chaotic with nearly random  motion,
and hence will not support the existence of the bar. Indeed earlier exploratory
work on this issue suggested that this is the case even for non-axisymmetric
halos with moderate constant density cores (El-Zant \& Hassler 1998).  This 
was  confirmed by self-consistent simulations performed by Ideta \& Hozumi
(2000) for systems with centrally-concentrated axisymmetric halos, but with
dominant disk contribution to the mass distribution of the inner regions.

It is the goal of this paper to investigate the prospects of bar survival for
different  halo shapes  in detail. We will do this by analyzing the response
of the orbital structure supported by the system to various degrees of halo
central concentration and asymmetry.
Since it has been shown (Dubinski 1994) that the settling of
a baryonic component in a triaxial halo can significantly reduce the initial
non-axisymmetry born of dissipationless collapse, we will only consider 
rather small departures from axisymmetry; deviations from unity in the
potential axis ratio of $10\%$ or less, which would induce ellipticities
in galactic disks of still  smaller magnitude and are, therefore, consistent
with observations of even present day galaxies (for a summary 
of observational results concerning halo shapes see, e.g., 
El-Zant \& Hassler 1998; Tremaine \& Ostriker 1999).

The stability of motion along trajectories in our models can be quantified by
a variety of methods developed in the dynamics of nonlinear systems (e.g.,
Hilborn 1994). In particular, we shall employ the method of Liapunov
exponents. This has the advantage of providing  a {\em timescale} that is
associated with an instability, i.e., an e-folding time, is much simpler to
implement with high accuracy than other methods (e.g., Laskar's frequency
analysis: Papaphilipou \& Laskar 1998) and is suited for far from
integrable systems where most
of the phase space is occupied by highly chaotic trajectories --- as will
turn out to be the case for some of our models.
It is readily generalized to higher dimensional,  
intrinsically time-dependent systems --- cases  where
other simple diagnostics, such as surface of sections are not applicable. 
It will be one of our goals to see how closely the Liapunov timescales 
correlate with their time-averaged density in configuration
(i.e., physical) space. Intuitively, it is expected that those trajectories 
that have small e-folding times will fill a large region of configuration
space,  generally not coincident with the bar density distribution. Therefore,
they are unlikely to serve as the main contributors in building a bar. If most,
or almost all, trajectories are of this type then one may plausibly conclude
that a bar is not sustainable. 
Sufficient conditions for a trajectory to belong to this group include
the following. The  volume a
trajectory occupies in configuration space
 exceeds substantially that of the bar at the radius where the trajectory starts.
The trajectory  spends most of the time outside the 
3-D figure of the bar, for example at vertical excursions larger than the 
bar's semi-minor axis. The trajectory is too isotropic in the plane, i.e. 
is too round to match the bar.

The methods which we employ in  evaluating the Liapunov exponents and 
the interpretation of their values are discussed in the next
section, the technical details are  deferred to the appendix. 
The spatial density distribution of the integrated trajectories  
will be examined  with the aid of the configuration space
grid described in Section~\ref{grid}. The system parameters of the galactic models and
the units used here can be found in Section~\ref{models}, and the initial
conditions and integrator employed are described in
Section~\ref{inicon}.  Section~\ref{stabs} contains the presentation 
of the main results of this paper. It
addresses the stability properties
of trajectories as well as their spatial distributions and the correspondence
of these two properties for models with various halo central concentration and
triaxiality. We will also be examining the vertical
stability of trajectories and examine its dependence, as well
as that of the general spatial structure,
 on the integration timescale. 
For all models we obtain the maximal Liapunov
exponents, which is faster to calculate than the whole set and is usually
indicative of the degree of instability of a given trajectory. For selected
models, we calculate the full set of Liapunov exponents. They are indicative of
conservation of invariants along trajectories. In the general time-dependent
potentials studied here, all the exponents can be non-zero. Finally, in
Section~\ref{endif}, we examine the correlation between the energy 
diffusion and the instability properties of the trajectories as 
determined by the values of the Liapunov exponents, for one of the models.
Our conclusions are discussed and summarized in Section~\ref{sec:conc}.    

\section{Liapunov exponents and stability of trajectories}
\label{Liapunov}

Motivated by the discussion of the previous section, 
we would like to estimate the fraction of initial conditions leading to
unstable (chaotic) trajectories and the associated timescale. For this purpose
we will calculate time-dependent Liapunov exponents for a large set of initial
conditions and display them on grayshade diagrams. In the appendix we
describe in a more formal manner how the exponents are obtained, here we
restrict ourselves to some general comments, helpful in the interpretation of
the results of the next section. 

Liapunov exponents compare the asymptotic rate at which the 
distance in the phase space, $\parallel \delta 
{\bf X}(t,{\bf X(0)}) \parallel$, between initially adjacent trajectories
starting with initial conditions around ${\bf X(0)}$ to the exponential 
(see for example Lichtenberg \& Lieberman 1995). They may therefore be 
written as
\begin{equation}
\lim_{t \rightarrow \infty}
\frac{1}{t} \log \frac{\parallel \delta {\bf X}(t,{\bf X(0)}) \parallel}{\parallel \delta X(0) 
    \parallel}.
\end{equation}
To each initial condition ${\bf X(0)}$ and 
perturbation $\delta$, there corresponds a Liapunov exponent, and to this
a characteristic exponential timescale (its inverse).  Thus a mapping is created
between the space of initial conditions and  the stability associated with
them.  

If $\parallel \delta {\bf X} (t) \parallel  \sim t$, 
as in the case, for example, of regular motion represented
by action angle variables (e.g., Binney \& Tremaine 1987) --- such that ${\bf
J}={\bf C}_1$ and ${\bf \Theta}= {\bf \omega} t + {\bf C_{2}}$,  with $\bf
C_1$ and $\bf C_2$ constant vectors --- the  exponents converge  to
zero as  $\sim \frac{\log t}{t}$.  If, on the other hand, 
$\parallel \delta {\bf X} (t) \parallel \sim e^{t}$, as in the case of
unstable systems,  then the maximal exponent tends to a  finite limit
(this limit can be shown to exist: see, for example, Eckmann \& Ruelle 1985).
One important difference between these two cases is that exponential instability
signals deviation {\em normal} to the trajectories and, therefore, implies
variation in the  action variables and not only the phase (angle) variables.
This can lead to time dependence.
For example, let us suppose one of the action variables
is the oscillations of the trajectory with respect to the $z$-axis. If this is
conserved, then trajectories starting near the $z=0$ plane will remain there.
If it is not conserved, then they will eventually venture out of that plane.
This may have important physical physical consequences, a system 
originally confined to a plane may become three dimensional.
How fast will such evolution in the statistical properties 
of trajectories take place will, 
in general (but not in all cases),  depend directly on the timescale of
the local instability.

For finite times, absolute distinction between regular and chaotic
orbits based on their Liapunov exponents is  impossible. It is
however  the exponential timescale  that is important in 
discerning the physical significance of the instability. 
It is sufficient, for our purposes, for the inverse of the Liapunov
exponent to be larger than a Hubble time for a 
trajectory to  be considered stable. 
Elementary estimates and test calculations show
that it requires  about $50,000$ Myr for the value of the maximal exponent of 
trajectories to reach the value of about $10^{-4}$ Myr$^{-1}$. Suppose, for
example, a flat rotation curve with a corresponding rotation period $T=\frac{2
\pi}{\omega} = 60$ Myr. In this case the divergence  between neighboring
circular orbits is  $\delta X \sim \omega t \sim 0.1 t$ and the exponents
after $50,000$ Myr is $\sim \ln 5,000/50,000 \sim 10^{-4}$.

The procedure of  integrating a trajectory for 
far more than the age of the Universe to test whether it is stable on a
smaller timescale can be rationalized in
the following manner. 
For one can show that in a model barred
potential most chaotic orbits  can be separated from regular ones by
inspection of their ``local'' Liapunov indicator $a_i= \ln
\frac{\xi_{i+1}}{\xi_i}$, where $\xi_i$ is an  initial infinitesimal
perturbation and $\xi_{i+1}$ is its consequent on a surface of section,
averaged over 10-20 consequents (Patsis et al. 1997). Note that except
for the averaging over  consequents and not over a fixed time interval,
which has no effect on the
result, the above scheme is equivalent to taking low $n$ sums of
Eq.~(\ref{eq:linf6}).  In this context then, one can divide each of our
trajectories into much smaller  segments, say
of 1 Gyr time-length each. Next, one can assume that each segment represent a
new trajectory with initial conditions corresponding to the terminal end of the
previous segment. Within this framework, the time-dependent Liapunov exponent
of this ``mega-trajectory'' can be considered an ensemble average over many
trajectories integrated over much smaller time. 

Because the exponential divergence can lead to mixing (as in a drop of ink
mixing in a glass of water), and because phase space density is conserved
along trajectories, an initial localized probability distribution will
diffuse and eventually fill the whole connected region with equal probability 
(see, e.g., Lichtenberg \& Lieberman 1995, and in the context of galactic dynamics
Kandrup 1998 and Merritt 1999). 
The rate at which this happens is characteristic of the
connected
region and is determined by its Kolmogorov entropy --- a measure of the 
rate at which the coarse-grained phase space volume occupied by such a group
of trajectory increases (see Appendix).
 When mixing is efficient one can replace
long time averages of quantities over a single trajectory by short time 
averages over ensembles of trajectories. 
In general, because of the
existence of semi-permeable phase space barriers and other complications (see,
e.g., Wiggins 1991), such assertions  are hard to prove in realistic dynamical
systems and may be difficult to ascertain numerically (because trajectories
may take very long  before they fill their allowed phase space region).
Nevertheless, test results by Kandrup \& Mahon (1994) suggest its basic
plausibility by showing that the local Liapunov indicators mentioned above,
averaged over ensembles of initial conditions in a connected  phase space
domain, approximate well the long time Liapunov exponent of a single trajectory
of the same region.

The configuration space density 
distribution calculated over long timescales can also
be thought to correspond to that of an ensemble of trajectories
integrated over a much shorter time.
Because of the qualifications  mentioned above however, it is still 
important to check the effects of the integration timescale 
on the distribution. This is especially true when examining transient effects,
like the time evolution of trajectories initially distributed  near a symmetry
plane into a three dimensional distribution. 
This will done in Section~\ref{shortev}.

Finally, a word about the calculation and interpretation of the full set 
of exponents. The exponential instability, when present, will lead any
perturbation to align itself with maximal direction of expansion in phase 
space; the most unstable direction.
 Thus
calculation of Liapunov exponents from a random perturbation will result in
the maximal one being found. In order to find the other exponents, one starts
from a set of orthogonal perturbations in phase space and re-orthogonalize them
at chosen intervals, to prevent them from re-aligning themselves again with
the direction of maximal expansion. Several methods have
been proposed for this procedure (e.g., Eckmann \& Ruelle 1985; Wolf et al.
1985). The method used here invokes the Gramm-Schmidt
algorithm and is described in the appendix. For systems of three degrees of
freedom six exponents are  found. Since  Hamiltonian systems have a skew
symmetric structure\footnote{Hamilton's equations are symmetric in the
variables except for the minus sign in one of them} they come in pairs of
positive and negative ones, meaning that to each direction of
expansion there is a corresponding  contraction (hence the
conservation of Poincare's invariants and phase space volume). It follows that
zero exponents come in pairs and correspond to integrals of motion.
In time-independent potentials there are
always two zero exponents, corresponding to energy conservation.
Other conserved quantities correspond to additional pairs of zero exponents.
Trajectories with zero exponents are regular, they conserve
three integrals of motion.

\section{Configuration space grid}
\label{grid}

For the purpose of  examining the configuration space distributions of 
our computed trajectories  we have devised a  $101 \times 10 \times
11$ cylindrical grid of bins $(IR,I\theta,IZ)$ 
defined as follows:
\begin{equation}
IR = INT \left[100 \frac{\log_{10}(R + 1)}{ \log_{10}(R_{max} + 1)} +1
   \right],      
\end{equation}
where $R_{max}$ is the maximum grid size taken to be double the bar's 
major axis and the last bin includes all radii lying beyond it,
\begin{equation}
IZ = INT \left[ 10 \frac{\log_{10} |z| \times 10 + 1)}
    {\log_{10}(z_{max}      \times 10 + 1)} \right],
\end{equation}
where $z_{max}$ is the maximum vertical grid size taken to be six times the 
vertical height of the bar, and again includes all the space above it.
Since the problem is symmetric with respect to the $x-y$-plane, only positive
values need to be recorded. Finally,
\begin{equation}
I\theta = INT \left[ \frac{\tan^{-1} |(y/x)| }{ 5 \pi} + 1 \right], 
\end{equation}
where $x$ and $y$ are Cartesian coordinates in a {\em frame rotating} with
the bar. Again, by virtue of symmetry with respect to the $x$ and $y$ axes only
one quadrant need be considered (hence the absolute value).

This frame provides high resolution in the center without being too coarse in 
the outer regions. This is necessary, as we will be considering trajectories 
ranging from $60$~pc to 6~kpc in radius, and starting from $20$~pc
in the vertical direction, but venturing 3-4~kpc out. The binning interval is
taken to be $0.05$ Myr. 

\section{Units and mass models}
\label{models}

We use the kpc as length unit and the Myr as the time unit. This fixes the
mass unit at about $2.2 \times 10^{11}~{\rm M_{\odot}}$, when the
gravitational constant is taken as unity.

We follow the traditional route of superposing distinct functional forms for
each of the components,  constructing our galactic models from separate
disk,  bar and  halo contributions. This procedure is an idealization, made possible by
the linearity of the Poisson equation and the corresponding additivity of the
forces due to different components. It allows one to distinguish the
different conceptual components. The equations of motion, however, are
nonlinear and the effects on the  motion due to the different components
cannot, in general, be  disentangled (since, obviously, the motion in a 
disk-halo system, for example, is not simply the motion in the disk potential
superposed onto the motion in the halo  potential). Our components are thus not
to be thought as approximating the density distributions arising
from different  kinematical components in a galaxy, which arise from the
nonlinear motion, by a linear superposition of uncoupled auxiliary components
(substitutes). The goal is to mimic in a reasonable way the galactic {\em mass
distribution}, especially those aspects most important to our analysis ---
radial density profile and symmetry properties.

A well known simple expression for a triaxial figure which has an
asymptotically flat rotation curve is 
\begin{equation}
\Phi_{H} = \frac{1}{2} v^{2}_{h0} \log \Bl R^{2}_{h0}+x^{2} + p y^{2} + q
z^{2} \Br , \label{loghalo}
\end{equation}
where  $v_{h0}$ is the maximum  rotation velocity  (invariably taken to
be $0.2 \sim 200$ km~s$^{-1}$) and $R_{h0}$ is the radius of a harmonic core. 
The parameters $p$ and $q$ are related
to the (ellipsoidal) equipotential axis ratios by $p=(a_h/b_h)^{2}$ and
$q=(a_h/c_h)^{2}$, with $a_h > b_h > c_h$ for a triaxial halo. 
In all our runs we fix $c_h/a_h=0.8$, while $b_h/a_h$ is varied in the 
range between $0.9$ and 1.

\begin{figure}[ht!!!!!!!!!!!!!!!!!!!!!!!!!]
\vbox to2.4in{\rule{0pt}{2.4in}}
\includegraphics{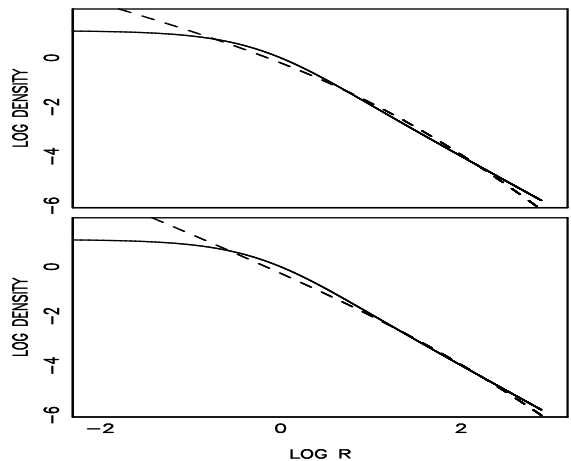}
\caption{Density of logarithmic potential (solid lines) as function of radius
(in units of the core radius $R_0$) as compared to those of closest
(by inspection) NFW model with central density increasing as $1/R$
(top), and more concentrated systems with density increasing as $1/R^{1.5}$
(bottom). \label{cnfw}}   
\end{figure}

The azimuthally-averaged density varies with 
radius as $\sim R^{-\alpha}$ with radius, with
$\alpha$ approaching 0 when  $R \ll R_{h0}$ and $2$ for $R \gg R_{h0}$.   It
is intermediate at radii of order $R_{h0}$. At these radii the density 
approximates that found at the inner and intermediate regions of 
halos identified in cosmological simulations of the cold dark matter structure 
formation scenario. Fig.~\ref{cnfw} 
compares the density profile of the logarithmic potential to the closest (by
inspection) models with inner density falling as $1/R$ and  $1/R^{1.5}$
--- corresponding to inner density profiles found by 
Navarro, Frenk \& White 1997 and  Moore \etal 1999, respectively.
In both these cases, the density falls off as $1/R^{3}$ in the outer regions.
 The cosmological halos have scalelength, determining the transition
from the inner profile to the outer one,  of
$3.7 R_{h0}$ and $10 R_{h0}$ respectively.

The main role of the disk component is to introduce strong asymmetry normal 
to its plane and to reduce it in that plane. Although an exponential disk is
most realistic, we have
decided to use the Miyamoto-Nagai (1975) model, because in
some cases we will be comparing our results with the self-consistent models of
Pfenniger (1984b), which had this form  for the disk potential.
It is also easily handled computationally with the force obtained in closed
form by taking the derivatives of the potential given by
\begin{equation}
\Phi_{D}= -  \frac{GM_{D}} {\sqrt{ x^{2}+y^{2}+
\Bl a_{d}+\sqrt{b_{d}^{2}+z^{2}} \Br^{2}} },
\label{tax:disk}
\end{equation}
where $GM_{D}$ is the mass of the disk, in units of $G=1$.
The parameters $a_{d}$ and $b_{d}$ determine the
central concentration and  the flatness of the disks. They are
taken to have values of $2.5$ and $0.5$ respectively. The disk mass is always
taken to be $GM_D=0.2$.

A rapidly rotating bar with a pattern speed $\Omega_b$ has been added in the
plane containing the halo minor axis. This plane also contains the disk when
it is present. The bar model used was a Ferrers ellipsoid of order two 
(Binney \& Tremaine 1987;
Pfenniger 1984a). The density of the Ferrers bar is constant along contours
given by 
\begin{equation} 
m^{2}=x^{2}/a_{b}^{2}+y^{2}/b_{b}^{2}+z^{2}/c_{b}^{2},
\end{equation} 
where $a_{b}>b_{b}>c_{b}$ are the semi-axes of the density distribution.
Inside the mass distribution ($m<1$) the density varies as 
\begin{equation}
\rho=\rho_{c} (1-m^{2})^{2},
\end{equation}
where the central density $\rho_{c}$ is determined by the total mass of the
bar $GM_{B}$. The density is zero for $m \geq 1$ .

The bar semi-axes are taken as $a_b=6$, $b_b=1.5$ and $c_b=0.5$, for comparison
with Pfenniger's (1984a,b) main model. 
The bar pattern speed was always chosen so that
the bar extends to corotation --- calculated while assuming an axisymmetric
halo and not including the bar's own contribution to the potential. Thus in
practice the bar ends inside corotation in accordance with empirical
requirement (e.g., Athanassoula 1992). In models where a disk was present, the
bar had $20\%$ of the disk mass within the circle 
tangent to the edge of its major axis.

\begin{figure}[ht!!!!!!!!!!!!!!!!!!!!!!!!!]
\vbox to5.4in{\rule{0pt}{5.4in}}
\includegraphics{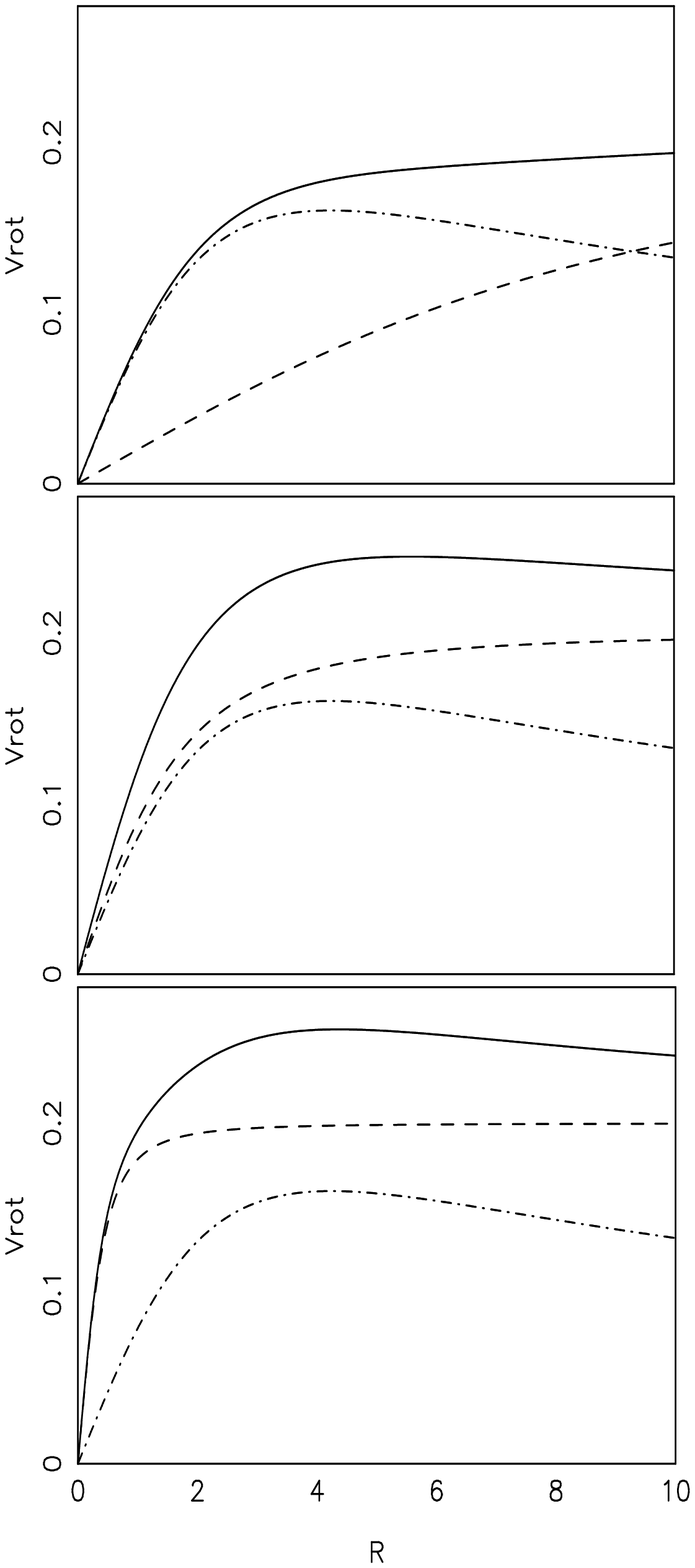}
\caption{Rotation curves of Model~1 (top), Model~2(center) and Model~3.
Dashed curves represent halo contributions, dashed dotted curves
are the disk rotation curves while the solid lines relate to the
total rotation curves. \label{rots}}   
\end{figure}

Models $1-3$ include all the components, the triaxial halo, bar and disk.
They are ordered in sequence of increasing halo concentration. Model~1
is a ``maximal disk'' model while Model~3 is halo dominated at all
radii. The rotation curves of these  three models are shown in 
Fig~\ref{rots}.
Models $4-6$ are without a disk. This makes it straightforward
to quantify the net contribution of the non-rotating 
non-axisymmetric perturbation ---  the (constant) 
 equipotential axis ratios
measured in the inertial frame
in the absence of the bar are simply those of the halo.
Model~7 has an axisymmetric halo and a bar
only, Model~8 corresponds to main model 
studied by Pfenniger (1984a) with a   
bar embedded in a disk without halo. Model~9
deals with triaxial halo only. The complete set of model parameters is given in
Table~\ref{table:models}.

\begin{deluxetable}{lccccccccc}
\tablecaption{Model Parameters}
\tablehead{
Model & $R_{h0}$ & $v_{h0}$ & $b_h/a_h$ & $GM_D$ & $a_d$ & $b_d$ & $GM_B$
   &$\Omega_b$ & Notes \nl }
\startdata

{\bf 1}    & 10.0   &  0.2   & 0.9   & 0.2   &  2.5    & 0.5  & 0.029 &
  0.031 & Triaxial halo with disk, bar  \nl 
{\bf 2}    & 2.0   &   0.2   & 0.9   & 0.2   &  2.5    & 0.5  & 0.029 & 0.041
   & Triaxial halo with disk, bar  \nl 
{\bf 3}    & 0.5   &   0.2   & 0.9   & 0.2   &  2.5    & 0.5  & 0.029 & 0.042
& Triaxial halo with disk, bar \nl 
{\bf 4}    & 0.5   &   0.2   & 0.9   & --   &  --    & --  & 0.03  & 0.033    
 & Triaxial halo, bar w/o disk \nl 
{\bf 5}    & 0.5   &   0.2   & 0.95   & --   &  --    & --  & 0.03  & 0.033 &
  Triaxial halo, bar w/o disk  \nl %
{\bf 6}    & 0.5   &   0.2   & 0.99   & --   &  --    & --  & 0.03  & 0.033  
   & Triaxial halo, bar w/o disk   \nl 
{\bf 7}    & 0.5   &   0.2   & 1.0   & --   &  --    & --  & 0.03  & 0.033     
&  Axisymm. halo, bar w/o disk \nl 
{\bf 8}    & ---   &   ---   & ---   & 1.0  &  3.0   & 1.0  & 0.1  & 0.052  
     &  Pfenniger model (disk, bar) \nl 
{\bf 9}    & 0.5   &   0.2   & 0.9   & --   &  --    & --  & --  &  --       
 & Triaxial halo only \nl 
%

\enddata
\label{table:models}
\end{deluxetable}

\section{Coordinate system, initial conditions and integrator}
\label{inicon}

We employ the coordinate system of Pfenniger (1984a), whereas the canonical
phase space variables are given by (x,y,z,X,Y,Z) and through these the 
Hamiltonian is defined as 
\begin{equation} 
H=\frac{1}{2} \Bl X^{2} + Y^{2} + Z^{2} \Br + \Phi(x,y,z) - \Omega_b \Bl x 
    Y -y X \Br,
\label{eq:ham}
\end{equation}
where the potential $\Phi$ is given by the sum of the contributions described 
in the previous section. In this system the spatial coordinates refer to 
a frame uniformly rotating with the bar with angular velocity $\Omega_b$,
 while the velocities are measured in the
inertial frame. The coordinates of the halo, assumed to be non-rotating, are
rotated back according to $x_h =x  \cos(\Omega_b t) - y \sin(\Omega_b t)$ and 
$y_h = x  \sin(\Omega_b t)  + y   \cos(\Omega_b t)$. Once calculated, the halo
force is then transformed back into the rotating frame in a similar manner.

Our goal is to investigate the stability of candidate trajectories which 
may support a bar embedded in different disk-halo configurations. For this purpose
the following initial conditions are appropriate. We start
slightly above the $x$-axis, directed along the bar major axis. The
initial amplitude of the $z$-offset is taken as 20~pc for runs, except
for those in Section~\ref{shortev} where the initial $z$ excursion is
increased to 200~pc. The velocities of the particles are taken normal to the
$xz$-plane: that is the only nonzero component is $Y$, also taken to be
positive. This means that our trajectories are symmetric with respect to the
$x$-axis and they are also, at least in the absence of  additional triaxial
perturbation, prograde. In the case where the halo is axisymmetric, these
initial conditions, within the limit of resolution in the initial $x$ and $Y$
values, produce all trajectories parented by the $x_1$ periodic orbits aligned
with the bar. These are known to be the building blocks of self-consistent
bars (e.g., Pfenniger 1984b, where they are termed B orbits).

The binning of the initial conditions is done on a linear equal-spacing grids
with 100 subdivisions in each of these coordinates. Thus in total we have, for
each model, 10,000 trajectories, with 100 of them starting from each $x$
position. The maximum $x$ position is taken to be the bar major axis (6~kpc)
and the maximal $Y$ velocity at each radius is $1.25$ times the local 
rotation speed assuming an axisymmetric halo (and neglecting the bar's own
contribution).

In some of the figures trajectories will be labeled sequentially
in the following manner. The trajectories with the lowest initial $x$ 
coordinate are taken to be the first hundred. They are ordered in ascending 
manner according to their value of $Y$. Thus the trajectory having the lowest 
rotation velocity is number one and that having the highest is number 100.
Next come the trajectories with the second lowest initial $x$ values, again
ranked in ascending order, according to their initial  $Y$ values. And so 
on in a way that the last hundred (those with rank 9,900 to 10,000)
trajectories  start with largest initial $x$ values in ascending order in
$Y$.

The integration is advanced using the variable order variable stepsize Adams 
method as implemented in the NAG routine D02CJF with a local (per timestep) 
tolerance of $10^{-10}$. Even at this low tolerance level there is no guarantee 
that the chaotic trajectories integrated are the actual trajectories of the system
from the given initial conditions. Indeed, varying the tolerance level gave usually 
different trajectories. This is because the problem is inherently unstable. 
Nevertheless, it was found that for the range of tolerance 
$10^{-8} - 10^{-14}$, trajectories were {\em qualitatively} similar, as 
evidenced by their occupation numbers on the grid of Section~\ref{grid}, for
example. In addition, when canonical conjugate initial conditions were chosen
for the tangent space vectors (i.e., vectors pointed towards pairs of position 
and velocity coordinates $(x,X)$, etc.), the Liapunov exponents came in
positive and negative pairs  to an accuracy of better than a few parts 
in $10^{8}$.
This is a good measure as to the accuracy of the calculation
since it implies  that the Poincare' invariants of all order
(e.g., Arnold 1989; Sussman, Wisdom \& Mayer 2001) 
are conserved to high accuracy and the symplectic
nature of the Hamiltonian  system is  thus conserved: 
despite the fact that the integrator is not symplectic by design, 
it is effectively so.

Except for Section~\ref{shortev}, where results are presented for trajectories 
integrated through a time interval of $10,000~{\rm Myr}$, all data relate to 
trajectories evolved for $50,000~{\rm Myr}$. Arguments rationalizing the choice
of this particular time interval are given in Section~\ref{Liapunov}.

\section{Stability of trajectories and configuration space
distribution} 
\label{stabs}

In this section we will analyze the stability properties of trajectories
and their relationship to trajectories distribution in the configuration
space. This will be achieved  by calculating the Liapunov exponents 
and by employing the cylindrical grid described in
Section~\ref{grid}. Our goal will be to examine the plausibility of 
self-consistent equilibria for barred galaxy models with different halo
structures --- differing from each other in their degree of  central
concentrations and triaxiality. In addition, we would like to know how 
tight is the correlation between the trajectories' distribution in the
configuration space and their stability properties.

\subsection{Maximal exponent and configuration space volume}
\label{max}

\subsubsection{The effect of central concentration}
\label{conc}

\begin{figure}[ht!!!!!!!!!!!!!!!!!!!!!!!!!]
\vbox to4.0in{\rule{0pt}{4.0in}}
\includegraphics{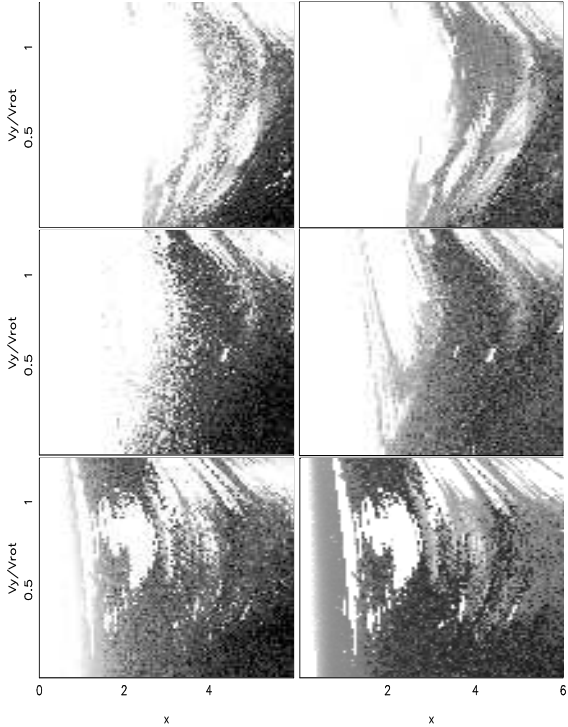}
\caption{Grayshade diagrams showing configuration space volume (left panels) and
maximal Liapunov exponents (right panels) for trajectories of Model~1 (top),
Model~2 (middle) and Model~3. The configuration space volume is defined as the
total volume  of the cells visited by a trajectory and is normalized in units
of the volume of the bar within the starting radius of the trajectory.
The scale is logarithmic (base ten) with the white background
corresponding to values of $-0.5$ and less, and the foreground to values of
$2$ (i.e., hundred times the volume of the bar at the initial radius) and more.
The exponents are rescaled, so that their inverse is given in units of
dynamical time (taken to be the local rotation period in the
azimuthally-averaged potential excluding the bar). The scale is
logarithmic.  The background corresponds to values of $-1.5$ or
less (i.e., exponential times of $\sim 32 \tau_D$ or more and the foreground
to values of $0.5$ or more.
\label{occucores}}   
\end{figure}

To get a measure of the configuration space volume occupied by a given
trajectory we calculate the total volume of the cells which it visits.
This volume is normalized in terms of the bar's volume 
within the trajectory starting point, i.e., the volume of the bar between
two planes normal to the major axis of the bar at the starting point
(and its reflection). The results for Models~$1-3$ are
shown on the left-hand panels in Fig.~\ref{occucores}, which depict a 
sequence of increasingly centrally-concentrated halos with core radii
decreasing from 10 kpc to 0.5 kpc. 
The models are presented in the form of grayshade
diagrams, with darker shading corresponding to larger relative volume occupied
by a trajectory. The positions on the diagrams correspond to the initial
conditions --- with the vertical axis rescaled in terms of
the local rotational speed at the initial  
 $x$ value labeled on the horizontal axis.  

\begin{figure}[ht!!!!!!!!!!!!!!!!!!!!!!!!!]
\vbox to3.0in{\rule{0pt}{3.0in}}
\includegraphics{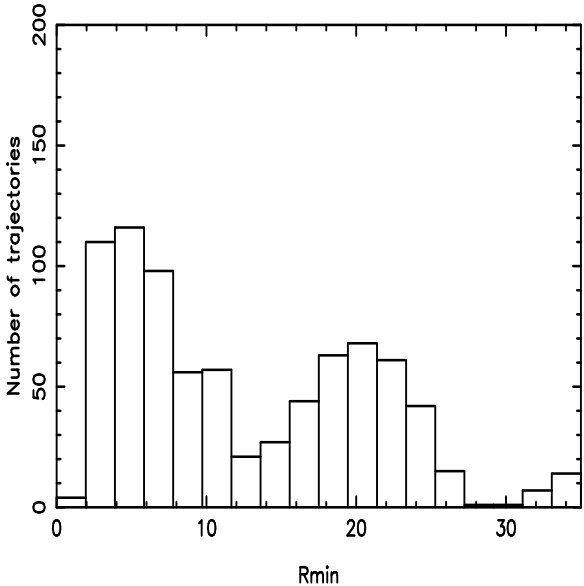}
\caption{Distribution of minimal radii of regular
trajectories (defined here as those having configuration
volume smaller than the bar's at their starting radius)
of Model~3  with initial coordinate $x \ge 3$.
\label{historegl} }   
\end{figure}

The darkest shading corresponds to trajectories occupying a configuration
space volume which is 100 or more in the normalized units (see above).
Clearly such trajectories (and all those that are represented by the darker
spots on this logarithmic scale) cannot contribute towards a self-consistent
bar --- by virtue of the fact that their  spatial distribution is far more extended
than the bar at their starting  radius. As can be seen, the fraction of such
trajectories increases significantly as the halo central
concentration is increased: being confined to the outer regions (where the
equidensity profiles of the  model are themselves more round) in
Model~1, but occupying the vast majority of all initial conditions in Model~3.
As we will see shortly, these trajectories are chaotic, with large
Liapunov exponents. 
And even though the volume of configuration space occupied by some 
of the chaotic trajectories in this model is  sometimes smaller 
than corresponding trajectories in the case of $R_{h0}=2$,
trajectories with volume significantly
larger than the bar are much more abundant.
Indeed, in this latter case, except for the ``island'' 
around $V_y/V_{rot} =0.75$
between initial condition with $1.5$ kpc  and $3$ kpc and the very 
inner region (where the halo potential is nearly harmonic) there are hardly
any regular trajectories that are elongated with the bar 
Moreover, many of the regular trajectories do not match
the bar density distribution, being too round by comparison.
This  is the case of the regular regions 
corresponding
to initial velocities $V_y/V_{rot} \ga 0.8$ and initial radii greater 
than 3 kpc. For 
as can be seen from Fig.~\ref{historegl}, where we plot the distribution 
of the minimum (cylindrical) radius visited by these trajectories,
 the majority of trajectories started beyond
$x=3$ have minimal radii that are greater
than the bars minor axis; they envelop the bar and thus cannot contribute
towards building a self-consistent model. Note that the trajectories which do
have minimal radii smaller than $1.5$ correspond to the isolated 
islands of stability   
corresponding to initial velocities $V_y/V_{rot} \la 0.8$.
One can, therefore, conclude that the sequence of Models $1-3$ represents
progressively more unstable bars which will tend to evolve quickly, since the
system does not support the kind of trajectories required to build such a bar
self-consistently. 

From the right hand-side panels of Fig.~\ref{occucores} one observes that most
of  the trajectories occupying large regions of the configuration space are
chaotic. Furthermore, comparison of left and right panels shows that there is
quite a tight correlation between the values of the Liapunov exponents and
the volume of configuration space the trajectories move in. This, of course,
should not come as a surprise.  For as explained in the introduction and
in Section~\ref{Liapunov}, regular orbits are confined by 
invariants which
characterize the regular motion while chaotic ones are under no such
constraints.  Trajectories with higher Liapunov exponents, in addition to
having smaller instability timescales, are also typically found in regions of
phase space away from regular trajectories and  barriers, such as cantori
which can prevent phase space transport (e.g., Wiggins 1991).  This  implies
that, in a given time, they are to diffuse in larger regions of phase space,
which is then reflected in their configurations space distribution.
Nevertheless, the degree of correlation is stark and shows the intimate
relation the question of stability of trajectories bears to that of
self-consistency of galactic models.

\begin{figure}[ht!!!!!!!!!!!!!!!!!!!!!!!!!]
\vbox to3.0in{\rule{0pt}{3.0in}}
\includegraphics{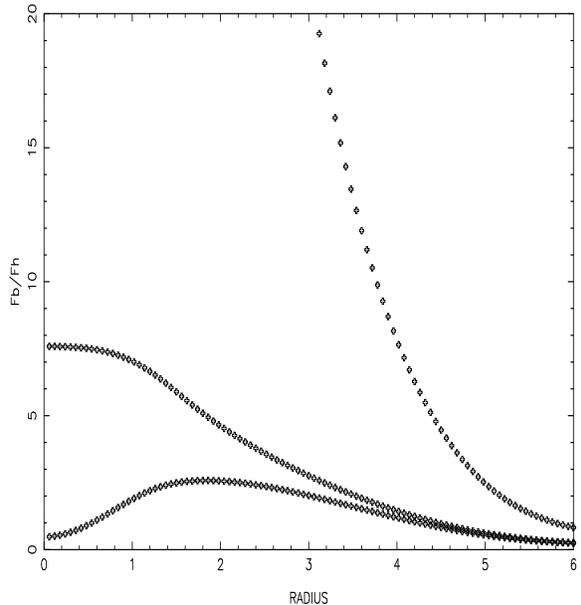}
\caption{Ratio of the average of the absolute value of the bar's
azimuthal force to that of the halo, for halo axis ratio
$b/a=0.9$ and for different core radii. The bar's relative
contribution is seen to be strongest when the halo has a large core
radius $R_0=10$, weakest for a centrally-concentrated halo $R_0=0.5$
and intermediate for $R_0=2$.
\label{dicores}}   
\end{figure}

There are two effects which may be leading to the increase in  the 
fraction of chaotic trajectories with the decrease in the halo core radius. 
The first, as discussed in the 
introduction, is the increased nonlinearity introduced by the 
centrally-concentrated mass distribution --- which leads to increased
coupling between the degrees of freedom. The second is the increased time
dependency in the force field. For, as can be seen from Fig.~\ref{dicores},
where we plot the ratios of the non-integrable components of the bar and halo
force fields, for the least centrally-concentrated halo ($R_{h0}=10$) it is
only at the edge of the bar that the average azimuthal force component 
of the halo becomes
comparable to that of the bar. In the case of $R_{h0}=0.5$, on the other hand,
this component is of the order of the corresponding bar component at all radii.
For mild halo triaxialities, therefore, it
is necessary that the halo be centrally-concentrated for the effect of time
dependency to become important. 

\subsubsection{The effect of halo triaxiality}
\label{triax}

\begin{figure}[ht!!!!!!!!!!!!!!!!!!!!!!!!!]
\vbox to4.0in{\rule{0pt}{4.0in}}
\includegraphics{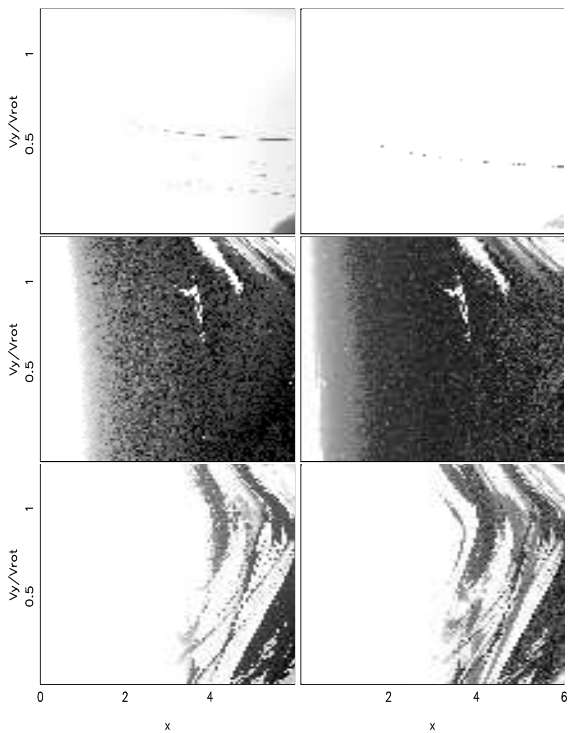}
\caption{Same as in Fig.~\ref{occucores} but for trajectories of
Model~9 (top), Model~4 (middle) and Model~8.
\label{selfcons}}   
\end{figure}

We now attempt to look in more detail into the origin of the widespread
chaotic behavior observed in the orbital structure of  barred 
systems with centrally-concentrated halo. 
To more easily identify the strength of the non-rotating 
non-axisymmetric contribution to the potential, we now remove 
the axisymmetric disk from our superposition of potentials. 
Removing the disk contribution will obviously increase the effective
non-axisymmetric perturbation in  the potential. 
This is expected to increase the region of instability even further. It is
apparent, however, that the dominant element producing unstable trajectories 
near the short axis plane of centrally-concentrated halos in our models
is not their own triaxiality but the presence of a rotating barred 
component. This can be seen from Fig.~\ref{selfcons}, where we show
the grayshade diagrams for a centrally-concentrated triaxial halo with
$b_h/a_h=0.9$, $c_h/a_h=0.8$. In the absence of the bar, trajectories are
regular and are confined in configuration space (Model~9, top panels). 
The addition of  the rotating
perturbation, however, modifies the situation dramatically (Model~4, center). 
For comparison, we
show the grayshade diagrams of  the model self-consistently constructed by
Pfenniger (1984b: Model~8). It is evident that the orbital 
structure, as manifested by
the configuration space distribution and Liapunov numbers, of
this model is different from that of models where a 
 centrally-concentrated triaxial halo
is present. This of course confirms our original inference that a 
self-consistent barred system is impossible to sustain inside a triaxial and
centrally-concentrated halo:  the trajectories it supports are too chaotic and
occupy too large a volume of the configuration space to represent a bar. 

\begin{figure}[ht!!!!!!!!!!!!!!!!!!!!!!!!!]
\vbox to4.2in{\rule{0pt}{4.2in}}
\includegraphics{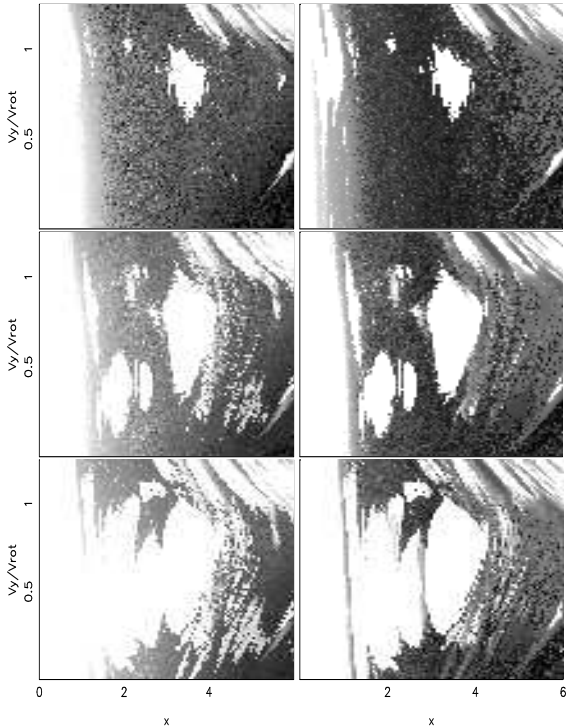}
\caption{Same as in Fig.~\ref{occucores} but for trajectories of Model~5 (top),
Model~6 (middle) and Model~7.
\label{moreax}}   
\end{figure}

In fact, as will be seen from Fig.~\ref{moreax} (Model~6, middle panels), 
even very small departures from axisymmetry on the part of the halo (as small
as $1\%$ in the potential axis ratio $b_h/a_h$) can lead to large connected
regions of initial conditions corresponding to  chaotic trajectories, with only 
islands of 
stability in their midst. Only in the limit of a perfectly axisymmetric halo 
models  is it possible to obtain connected regions of regular trajectories at 
most radii. In that case (bottom panels), significant fraction
of the initial conditions are occupied by trajectories that support the bar
--- the white regions that appear for most  initial
$x$ up to about 4~kpc represent trajectories that are mostly 
parented by stable  periodic $x_1$ orbits,  and are thus aligned with the bar.
\footnote{The periodic
$x_1$ orbits survive,  in distorted but stable form, for mild halo
triaxialities ($b_h/a_h \ga 0.92$). For potential axis ratios of $0.9$ or
smaller, they are (at least mostly) unstable. Detailed examination of the
existence and stability  of these periodic orbits in time-dependent potentials 
will be presented elsewhere.}
The existence of these trajectories is necessary for the construction
of barred equilibria. It is consistent with  numerical
simulations where long lived bars embedded in centrally-concentrated
axisymmetric halos are observed  (Ideta \& Hozumi 2000;
Debattista \& Sellwood 2000; Athanassoula \& Misiriotis 2002).
These may slowdown in time but do not dissolve.
In some of these simulations rings surrounding the 
bars are also found. These are possibly related to the trajectories
represented by the white strips on the upper right corner of the 
diagrams in the bottom panels of Fig.~\ref{moreax}. These regions 
correspond to nearly round trajectories which, at the outer edge of the bar
and slightly beyond, are elongated in the direction perpendicular 
to the bar --- as is observed in the simulations.

The fraction of bar supporting regular  trajectories decreases
rapidly as small non-axisymmetric perturbations are 
applied.
In the 
case of Model~6 orbits elongated along the bar still exist at most radii.
However they occupy  a much smaller fraction of the initial conditions
than in Model~7. The range of shapes they come in is therefore narrower, 
which implies that it is less probable that among them will be found those 
that match the bar's asymmetry (i.e., are not too fat). Indeed, we find by
inspection that a large fraction of the available trajectories that
are elongated along the bar do not match its asymmetry. 
Further departures
from axisymmetry render the bar supporting orbits 
exceedingly rare. In Model~5 (Fig.~\ref{moreax}, top panels) 
they are found at a small
starting range in $x$. In model~4, they are virtually non-existent.

The above implies  that our  axisymmetric model is 
{\em structurally unstable}
 --- since any small perturbation renders the once connected regular regions 
into disjoint sets. 
The replacement of regular trajectories parented by the $x_1$ orbits
by chaotic ones in turn implies that self-consistent bars are unlikely under
these circumstances. Because the perturbations required are very small, 
axisymmetric models of barred galaxies with centrally-concentrated halos 
must then be considered non-generic.

\subsection{Vertical stability of trajectories}
\label{vert}

\begin{figure}[ht!!!!!!!!!!!!!!!!!!!!!!!!!]
\vbox to4.2in{\rule{0pt}{4.2in}}
\includegraphics{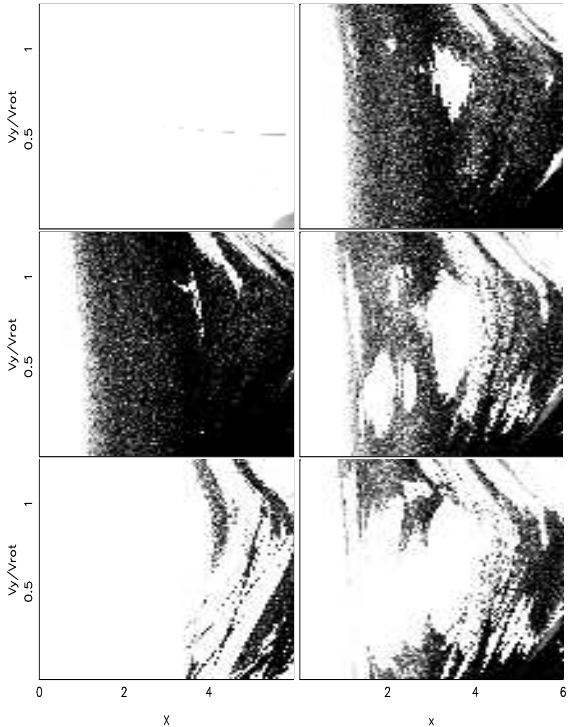}
\caption{Grayshade diagrams showing the fraction of time trajectories spend
at cells comprising vertical coordinates larger than the bar
minor axis. Left panel: Model~9 (top), Model~4 (middle) and
Model~8 (as in panels of Fig~\ref{selfcons}). Right panel: Model~5 (top),
Model~6 (middle) and Model~7 (as in panels of Fig~\ref{moreax}).
The shades are on a linear scale whereas the foreground corresponds
to trajectories spending $75\%$ of their time or longer in the
aforementioned cells and the background corresponds to trajectories
spending $0.075 \%$ of the integration time in such cells.
\label{zgfigs}}   
\end{figure}

The conclusion that systems with centrally-concentrated triaxial halos are
unlikely supporters of bars is impressed further when one looks at the
vertical stability of the trajectories under consideration. In
Fig.~\ref{zgfigs} we show, for some of our models, grayshade diagrams
representing the fraction of time trajectories spend in cells that correspond
to vertical excursions greater than the semi-minor of the bar. It is 
evident that for models with even minor triaxiality this fraction can exceed
one-half of the time for a large number of trajectories.  Note that this is
necessarily an {\em underestimate} --- the bar's extension is, except at the
origin, always less than the minor axis value.

Comparison of the structure of the
grayshade diagrams in Fig.~\ref{zgfigs} with those in Fig.~\ref{selfcons} and
Fig.~\ref{moreax} reveals that the fraction of time a trajectory spends
having large absolute values for its vertical coordinate correlates
well with the values of the Liapunov exponents. In other words, trajectories
which spend the largest fraction of time at large $z$ have largest Liapunov
exponents. This in turn correlated with the total configuration volume they
occupy. Thus vertical stability is a simple 
diagnostic of chaotic behavior: almost all regular trajectories conserve 
their initially small amplitudes of $z$ oscillations (the exceptions
are those near bifurcations leading to three dimensional regular orbits). 
It is also 
a sufficient criterion that orbit densities do not match that of the bar, 
being too extended in the vertical direction.

We find that the maximal 
vertical excursion of chaotic  trajectories of some
of our models can be quite large --- of the order of 10~kpc. To illustrate
this, we have arranged the trajectories in ascending order according to their
starting spatial and velocity coordinates (see Section~\ref{Liapunov}) and
plotted the logarithm of their maximal vertical excursion within the first
$10,000$ Myr. The results are shown in Fig.~\ref{zline}.

\subsection{Evolution on shorter timescale and the
instability in- and out-of-the plane}
\label{shortev}

\begin{figure}[ht!!!!!!!!!!!!!!!!!!!!!!!!!]
\vbox to4.1in{\rule{0pt}{4.1in}}
\includegraphics{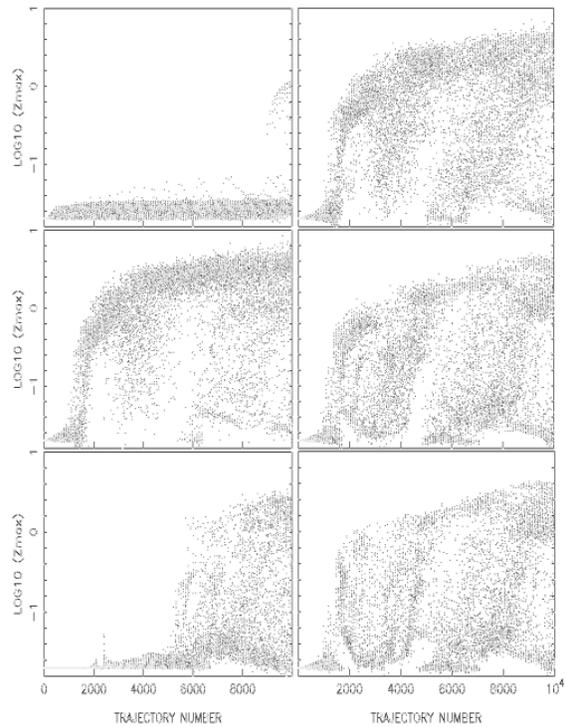}
\caption{Maximal vertical excursion within $10,000$ Myr. The panels correspond to those
of models shown in~Fig.~\ref{zgfigs}.  The trajectories are arranged in
ascending order such that those at smaller initial radii have strictly smaller
number than those starting at a larger radius. At each radius the rank
progressively increases with initial normal velocity
(cf., Section~\ref{inicon}).
\label{zline}}   
\end{figure}

Some trajectories which have large Liapunov exponents, and according to the
diagrams of Fig.~\ref{zgfigs} spend a large fraction of their time at $z$
coordinates larger than the bar's vertical extension, nevertheless appear to 
have small maximal vertical excursion in Fig.~\ref{zline}. This happens because
many of these trajectories do not, for the small initial $z$ amplitude we
used, reach their maximal $z$ extension in $10,000$ Myr.  We had chosen the
small initial $z$ amplitudes in order to detect trajectories that may be
vertically unstable even if their maximal excursions do not exceed a few
hundred pc. In realistic situations however, vertical motions would be
sufficient so that stars 
already have $z$ amplitudes of such magnitude. 

\begin{figure}[ht!!!!!!!!!!!!!!!!!!!!!!!!!]
\vbox to4.1in{\rule{0pt}{4.1in}}
\includegraphics{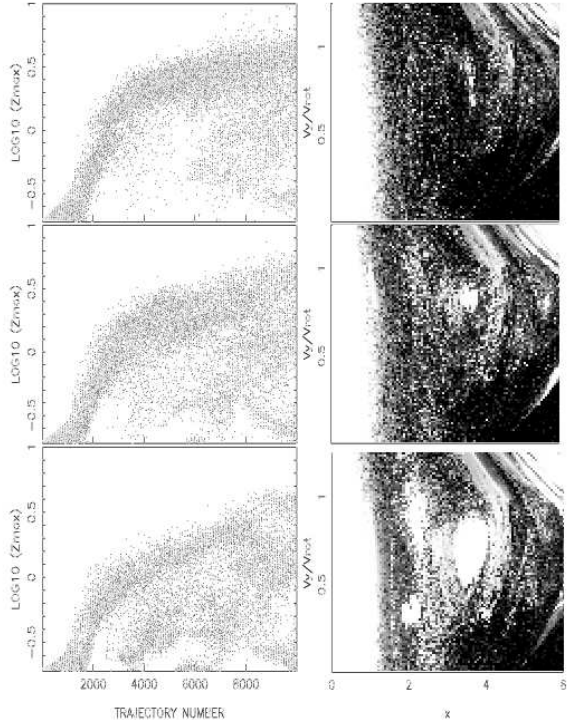}
\caption{Maximum vertical excursions (left panels) and grayshade diagrams showing the
time fraction spent  at configuration space cells comprising vertical
coordinates  larger than the bar minor axis for Model~4 (top), Model~5
(middle) and Model~6 (with the same scaling as in Fig.~\ref{zgfigs}.
Trajectories are integrated for  $10,000$ Myr starting with a vertical
amplitude of $0.2$~kpc (instead of the canonical $0.02$~kpc used in the other
runs).
\label{zlines}}   
\end{figure}

We have rerun some models with the larger initial  $z$ amplitude of $0.2$. The
results, for Models~4,~5 and 6  are shown in Fig.~\ref{zlines}.
One finds that many trajectories, that apparently were stable in the $z$ 
direction 
over the Hubble time, are now unstable. 
Fig.~\ref{zlines} (right 
panels) also displays diagrams, similar to those in Fig.~\ref{zgfigs}, 
representing
the fraction of time trajectories spend in cells comprising vertical
coordinates larger than that of the bar vertical extension. This
fraction of time now has increased significantly by starting our trajectories
at a larger $z$ coordinate value, albeit one that still is significantly
smaller than the bar's semi-minor axis. This suggests that the vertical instability
is important in determining the orbital properties of the system, 
transforming trajectories that are stable in the plane into unstable ones,
but that it manifests itself over relatively long time-scale, 
especially when the initial perturbation
is small.

Trajectories  having  large vertical extensions 
and spending most of their time 
with absolute values of their $z$ coordinates 
larger than the bar's semi-minor axis cannot
contribute towards a self consistent barred galaxy model. 
This being the case for most trajectories of Models~4,5 and~6, 
makes it apparent
that the bar would not survive in such models for a Hubble time 
--- which places
an upper limit on the dissolution time of any barred system embedded in 
mildly triaxial centrally concentrated halo. A lower limit, which should be 
of the order of the Liapunov timescale of the chaotic trajectories, is
of the order of a few dynamical times. This should be the time-scale 
for bar dissolution in the most chaotic systems (e.g., Model~4) where 
hardly any regular trajectories exist. In such cases, the vertical
instability is unlikely to be of central importance in the evolution
of the system --- since the bar would probably  dissolve in the plane before
its  presence becomes felt.

Fig.~\ref{shorts} shows the maximal Liapunov exponents and the configuration
space volume for Models~5, 6 and 7 for trajectories integrated through this
smaller period of time and starting with larger $z$ amplitude (note the 
difference in grayscales of Liapunov exponents). Again one observes a clear 
similarity between the distribution of values of the exponents
and the occupied configuration space volume.  The figures are qualitatively
similar to the ones integrated for $50,000$ Myr from  $z_{in}=0.02$
(cf. Fig.~\ref{moreax}). Quantitatively there are some differences. The
extent of the shaded regions has increased. Unstable trajectories are now more
abundant. This is again an effect of the larger initial vertical amplitudes
used here,  causing chaotic trajectories to replace regular ones. However the
most unstable trajectories now have somewhat smaller configuration space
volume than previously. This suggests that trajectories continue to explore
new regions of  phase space as the system evolves in time. It is not
surprising, given the time-dependent nature of the potential (meaning an
orbits phase  available space is not necessarily bounded) and the possibility 
of the presence of ``sticky'' chaotic trajectories that take long times
to achieve an invariant distribution even for the  time-independent case
(e.g., Siopis \& Kandrup 2002).  

\begin{figure}[ht!!!!!!!!!!!!!!!!!!!!!!!!!]
\vbox to4.1in{\rule{0pt}{4.1in}}
\includegraphics{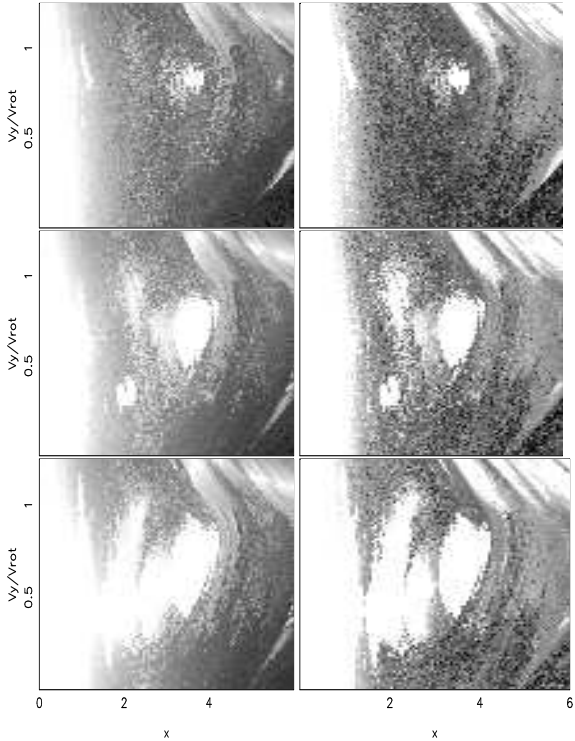}
\caption{Same as in Fig.~\ref{moreax} but for trajectories integrated for $10,000$ Myr
and with initial vertical condition $z=0.2$~kpc (instead of the canonical
$0.02$~kpc used
\label{shorts} in the runs in that figure). Note that, to adjust for the
significantly shorter integration time, which implies that regular
trajectories would have an exponent that is  proportionally larger, the
background for the grayshades representing these quantities  is rescaled to
-0.9 (instead of -1.5 as in Fig.~\ref{moreax}).
\label{shorts}}   
\end{figure} 

Indeed simple inspection of the spatial distribution of trajectories  showed
that both these effects are at work; some trajectories keep  on moving towards
larger and larger radii and vertical excursions,  while others stay within a
bounded region but fill larger and larger  volumes within that region. In the
latter case this mainly consisted of trajectories with a hollow configuration
space distribution, which  was being filled up as the integration proceeds. There
are also some trajectories whose vertical extension is only significant after
10 Gyr. None of the trajectories examined however had a spatial configuration
that did not match the bar over 50 Gyr but did match the bar over $10$ Gyr.  
In fact, the vast majority already showed
a spatial  structure in the plane which is very different from that of  the
bar within 500 Myr or so --- their shapes  being  too isotropic, even
though they were launched from initial  conditions corresponding to thin
bar-aligned orbits in the case when  the halo is axisymmetric. 
Vertical instability generally took longer to develop, but was far faster,
as expected, when the initial vertical perturbation was larger.

\begin{figure}[ht!!!!!!!!!!!!!!!!!!!!!!!!!]
\vbox to3.0in{\rule{0pt}{3.0in}}
\includegraphics{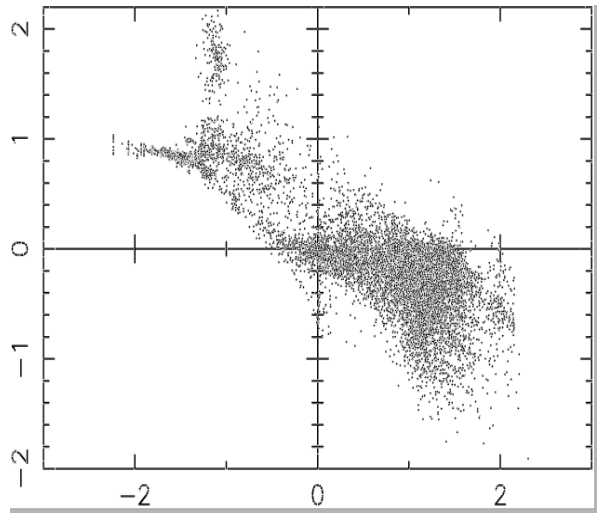}
\caption{Comparison of the configuration space volume for trajectories
integrated for $10000$ Myr (and stating AT $z=0.2$) and ones
integrated for $50000$ Myr (and starting at $z=0.02$). The
the horizontal axis represent values of the (base ten) logarithm of
the configuration space volume of the latter trajectories,
while the vertical axis represents logarithms (again base ten)
of the ratio of the two volumes for the different
trajectories of Model~5 (i.e., $vol(10000,i)/vol(50000,i), i=1, 10000$).
\label{relvol}}   
\end{figure}

Furthermore,
trajectories that occupied a relatively (compared to the bar) large
configuration space  over the longer times integration still 
do over the span of $10$ Gyr.    This can be seen from
Fig.~\ref{relvol} where we plot the  relative volumes of the trajectories of
Model~6 for the two cases against the volumes of the case where the
trajectories  are integrated for $50$ Gyr. As expected, trajectories  which, 
for the case of long integration from small initial $z$, had small
phase space volumes
usually acquire a much larger one.
This is due to the enhanced  $z$ instability which further destroys 
the regular regions. 
On the other hand, trajectories which had large 
configuration space volumes compared to that of the bar and had pronounced
$z$ excursion, even when starting from $z=0.02$, now have smaller volume. 
Nevertheless, this 
is still much larger  than the that of the bar --- again confirming that 
they are unlikely to contribute towards a self-consistent model on the
timescale of $10,000$ Myr.

\subsection{Full set of Liapunov exponents and distribution of exponential 
    times}
\label{fulset}

\begin{figure}[ht!!!!!!!!!!!!!!!!!!!!!!!!!]
\vbox to4.1in{\rule{0pt}{4.1in}}
\includegraphics{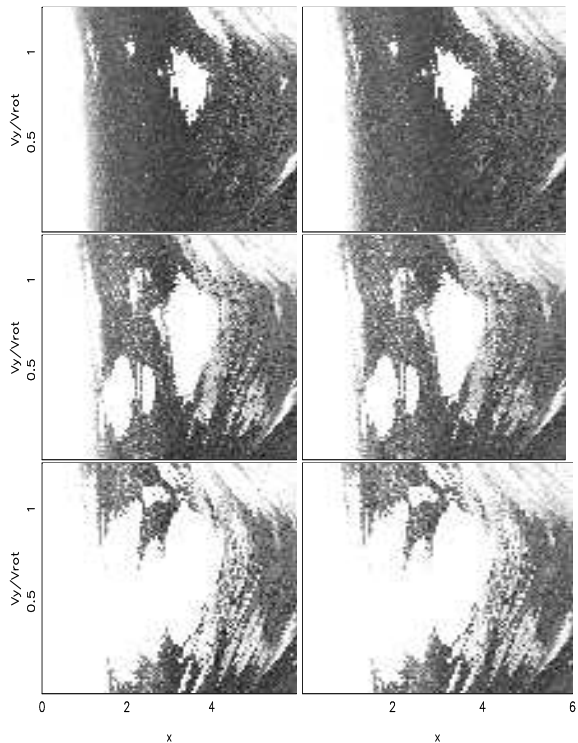}
\caption{Grayshade diagrams for the middle (left panels) and smallest Liapunov
exponents for Model~5 (top), Model~6 (middle) and Model~7. On the (base ten)
logarithmic scale the middle exponent background value is $-1.5$ and
foreground is $0.2$ while for the smallest exponent the background is scaled
to $-2$ and  foreground to $0$.
\label{comfig}}   
\end{figure}

Because it is CPU-time consuming to calculate the full set 
of Liapunov exponents at high resolution in the initial conditions, 
we performed this task for a selected set of models. The values of
these exponents provide information on whether trajectories, though being
chaotic, may conserve (even if approximately) some quantities with time.
Fig.~\ref{comfig} shows grayshade diagrams  involving  the
second and third Liapunov exponents of trajectories in the three models, 5, 6
and 7, whose configuration space volume and maximal exponents are shown in 
Fig.~\ref{moreax}. The agreement between the 
configuration volume of trajectories and the values of their exponents
is even  more precise   
than in the case of the maximal exponent --- most
discrepancies, even though minor originally, having now disappeared.
The additional constraints provided by low values in the two smaller 
Liapunov exponents account for the remaining regions of initial conditions 
corresponding to small configuration space volume despite having large values
of  the maximal exponent.

The sum of the positive exponents for a given trajectory characterizes its
Kolmogorov entropy (see Appendix). For a group of nearby trajectories in
connected phase space region it measures the rate of increase of the 
coarse-grained volume they occupy, and thus of the evolution of their
statistical distribution. The inverse of this entropy (strictly speaking
multiplied by a numerical factor of order unity that is ignored here) defines a
timescale  for such evolution.

\begin{figure}[ht!!!!!!!!!!!!!!!!!!!!!!!!!]
\vbox to4.4in{\rule{0pt}{4.4in}}
\includegraphics{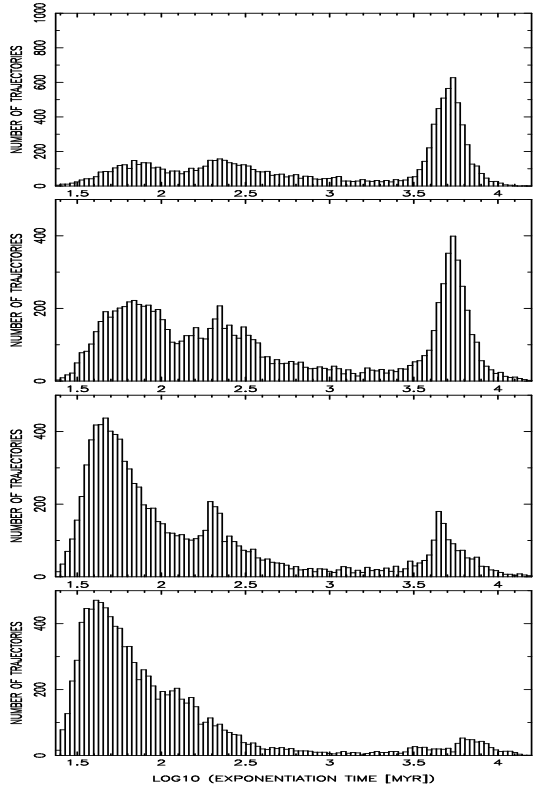}
\caption{Histograms showing the distribution of the exponentiation times (taken to be
the inverse of the  Kolmogorov-Sinai entropy) for trajectories of (from top to
bottom) models~7, 6, 5 and 4.
\label{histos}}   
\end{figure}

In Fig.~\ref{histos} we display histograms showing the distribution of  the
inverse of the Kolmogorov entropy for trajectories of Models~7, 6, 5 and 4. 
As can be seen, for trajectories of axisymmetric Model~7, the corresponding
exponential times are mostly very large, of the order of the Hubble time.
These correspond mainly to trajectories trapped around the $x_1$ periodic orbit
family aligned with the bar (the large connected white patch in the grayshade
diagrams in the bottom  panel of Fig.~\ref{moreax}).
Moreover, the more unstable trajectories, 
with smaller exponential timescales, are of roughly equal numbers over a
large range of values; the distribution is relatively flat, being only
slightly bimodal on far left of the figure. These are 
exponents of  various higher order regular or trapped chaotic
trajectories ---  i.e., ones existing in
regions of phase space where mainly regular orbits dominate.  The final peak
corresponds to trajectories starting from initial conditions in the outer
region of the system  (cf., Fig.~\ref{moreax} bottom panel),
 which correspond to
regions of phase space that are dominated by chaotic trajectories. 

As the non-axisymmetric perturbation is increased, regular trajectories
aligned with the bar progressively become less abundant. Initially there is an
equal increase in the number of trapped trajectories (second peak on left for
histogram of Model~6) and highly chaotic ones (first peak). Eventually,
however, the increase in the latter occurs at the expense of the former
(Model~5).  Finally, the fraction of both the trapped and regular
trajectories becomes very small (Model~4). 
Here most trajectories are highly
chaotic, part of the same connected ``stochastic  sea,''  with
exponentiation timescale varying relatively little. This is expected
since, for Model~4 (cf. Fig.~\ref{selfcons}, middle panel), 
except for small
regular regions, trajectories  form a connected region of
initial conditions that have similar Liapunov exponents and configuration space
volumes.

For  Model~4, most of  the  remaining regular trajectories are 
parented by surviving higher order periodic orbit families of the resonant
zones (the white strips in Fig.~\ref{selfcons} middle panels). 
None are parented by the $x_1$ family of the unperturbed axisymmetric 
halo case. Almost certainly,
the absence of a significant fraction of regular orbits, more importantly
the total absence of {\em any} regular orbits aligned with  the bar, ensures
that such a bar cannot be built self-consistently. 
For in the case of a time-independent potential, for example,
any set of  initial conditions starting in  a connected chaotic region
will evolve to an invariant distribution, with every trajectory having
the same time-averaged phase space distribution
in  as all other trajectories. The projection
onto configuration space of such a distribution would be too
isotropic  to match a bar.
In the case of a time-dependent system,
the region in which trajectories move is not necessarily bounded by
a zero velocity curve,
and therefore some trajectories can explore even larger volumes
as their energies change --- an effect that is likely to make the 
discrepancy more pronounced.

\section{Energy decorrelation}
\label{endif}

\begin{figure}[ht!!!!!!!!!!!!!!!!!!!!!!!!!]
\vbox to3.4in{\rule{0pt}{3.4in}}
\includegraphics{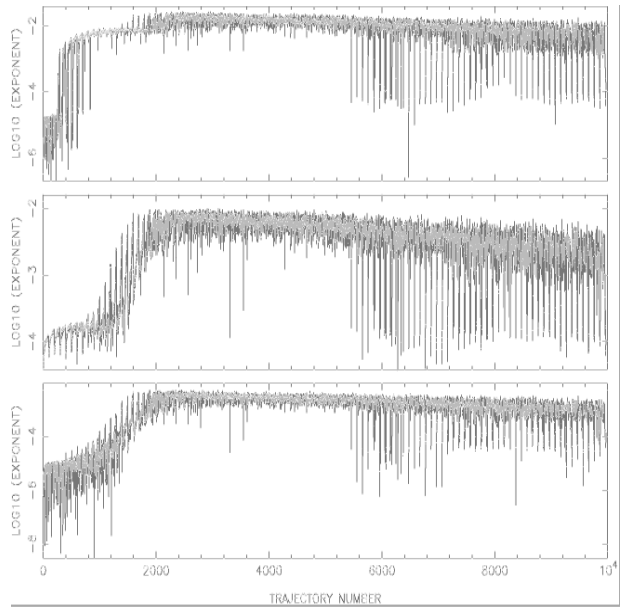}
\caption{Liapunov exponents of Model~5 for trajectories arranged sequentially
such that the first hundred correspond to initial conditions at the first
position bin and progressively larger normal velocities
(cf., Section~\ref{inicon}).
\label{liaplot3}}   
\end{figure}

For trajectories with large maximal exponent, the other two positive exponents
are proportional to it with a nearly constant of proportionality. 
This is not always true, however, of trajectories with smaller maximal 
exponent. The latter tend to have relatively even smaller corresponding values
for the two smaller exponents. This is illustrated in Fig~\ref{liaplot3}
where we plot the Liapunov exponents of trajectories of Model~4. It is
especially clear for trajectories starting in the inner regions
(those with small trajectory numbers). 
It turns out that such
trajectories also conserve their energy well.

Energy, always an invariant quantity in a time-independent potential, is
allowed to vary in  the time-dependent models described here. Corresponding to
energy  conservation is a pair of zero Liapunov exponents. 
Therefore, trajectories that 
conserve energy well will have at least one small exponent. In order to test
energy (i.e., the Hamiltonian of Eq.~[\ref{eq:ham}]) conservation, we
calculated the correlator    $C_i = \frac{< E(t) E(t + \tau i) >}{<E(t)>
<E(t+ \tau)>}$, where $\tau$ is a time interval, taken as 500 Myr, 
 and $i$ is an integer. The quantity $C_i$ as  defined above is zero
if there is complete loss of memory in energy  over a timescale $\tau i$, and
one if there is no loss whatsoever, ---  for example, if energy is conserved along a
trajectory. It can also be seen as an angle between a normalized unit vector,
consisting of the values of the energy at intervals $\tau i$, and another
``delayed'' vector with corresponding  points delayed from the first vector
also by an interval $\tau i$.

\begin{figure}[ht!!!!!!!!!!!!!!!!!!!!!!!!!]
\vbox to3.4in{\rule{0pt}{3.4in}}
\includegraphics{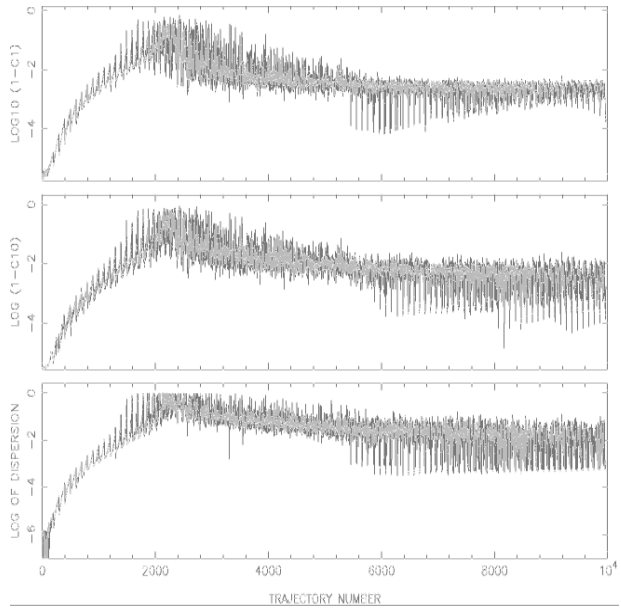}
\caption{Energy correlators $C_i = 1 - \frac{< E(t) E(t + \tau i) >}{<E(t)> <E(t+
\tau)>}$ averaged over $50,000$ Myr with interval $\tau i$ and $\tau = 500$
MYr (Top: $i=1$, Middle: $i=10$), and relative energy dispersions (bottom) for
Model~5. Trajectories are arranged sequentially
such that the first hundred correspond to initial conditions at the first
position bin and progressively larger normal velocities
(cf., Section~\ref{inicon}).
\label{disp4}}   
\end{figure}

In Fig.~\ref{disp4} we plot $\log_{10} (1-C_i)$ for $i=1, 10$, as well as the 
relative dispersion in energy along our enumerated trajectories of Model~4.
As expected, in the region where two of the Liapunov exponents have very
small values, energy correlation is large (that is near 1) and the  dispersion
is small. What is somewhat surprising is that, for larger radii, the energies
of the trajectories at different times  appear to be again correlated --- this
despite the fact that these trajectories
belong to large connected regions with high values of the exponents
(cf. Section~\ref{triax}). 

\begin{figure}[ht!!!!!!!!!!!!!!!!!!!!!!!!!]
\vbox to3.0in{\rule{0pt}{3.0in}}
\includegraphics{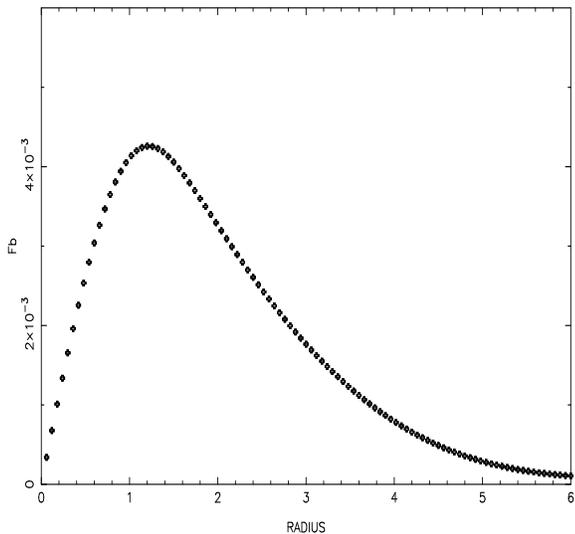}
\caption{Average (of absolute value of) azimuthal bar force as a function of radius.
\label{fabaraz}}   
\end{figure}

The variation of the dispersion  and the quantities $\log_{10} (1-C_i)$  as a
function of radius appear to follow closely the strength of the bar
perturbation (Fig.~\ref{fabaraz}).  This
trend does not seem to be substantially altered by changing the 
asymmetry of the halo (we have checked this by comparing  the results
presented here with analogous ones for Models~5 and~6). Thus, for any 
non-axisymmetric perturbation between one and ten percent in the
nonrotating potential, energy conservation of trajectories 
starting at radii
where the bar asymmetry is maximal is affected considerably.
For other trajectories, energy conservation is not significantly
affected by such 
small perturbations. One can perhaps say that the value of their 
Hamiltonian is stable along their trajectories.

It is interesting that
a time-dependent perturbation can have such a large effect  on the stability
properties of some trajectories and the manner in which their  configuration
space distribution behaves, but causes very little change in energy. 
It illustrates the fragility of orbital structure of 
the unperturbed axisymmetric system.

\section{Conclusions}
\label{sec:conc}

In this paper we have analyzed the stability and configuration space 
properties of
trajectories starting near the plane containing  a bar in galaxies with
different dark matter halo distributions. In particular, we examine the
variation of these properties  for different concentrations
(Section~\ref{conc})  and  triaxialities (Section~\ref{triax}). This is done
by calculating the Liapunov exponents, looking at the time-averaged spatial
density distribution (Section~\ref{stabs}) and the vertical 
 stability properties (Sections~\ref{vert}) of
samples of  $10,000$ trajectories  starting near the plane containing the bar
in each model.   As a reference we have  compared these properties with those
of the relatively stable model that has been built self-consistently by
Pfenniger (1984a,b). For some of the models we also calculate the full set of
Liapunov exponents and look at the distribution of  exponential timescales
inferred from the Kolmogorov entropy (Section~\ref{fulset}). In general, there 
is remarkable correlation between the local stability properties, 
as characterized
by the exponents, and the the spatial distribution of trajectories, as 
quantified, for example, by the configuration space volume they fill --- 
with the more unstable
trajectories occupying much larger volume than more chaotic ones.

For models with centrally-concentrated halos, we find a large fraction of 
trajectories to be chaotic. These fill  a large (compared to the bar's) volume
of the configuration space and spend most of the time at vertical
extension greater than the maximum thickness of the bar. They are, therefore,
unlikely to contribute towards a self-consistent  model. In addition many of 
the trajectories that remain regular have minimum extensions in the plane 
larger than the 
bar's middle axis. They also cannot contribute towards a self-consistent model
(Section~\ref{conc}).

The above is even  true to some extent in the case when the halo 
is centrally-concentrated but axisymmetric. 
Still, in this case, there remains large connected 
sets of initial conditions
corresponding to regular orbits which are aligned with the bar and 
could contribute to a self-consistent model.
When, however, the halo is triaxial, chaotic  trajectories become a dominant
majority --- even very small halo triaxiality (a $1\%$ perturbation in the 
potential axis ratio) causes the chaotic regions to become connected and
regular phase space  regions to become islands of stability
(Section~\ref{triax}).  Such  models are 
  unlikely candidates for steady state self-consistent
solutions.

There are several reasons why centrally-concentrated halos, especially triaxial
ones, contribute to the chaotic behavior displayed from most initial conditions.
A centrally-concentrated mass distribution has solutions for the Poisson
equation that is far from quadratic, in the sense that an expansion of the
potential power series has large terms beyond the quadratic. This produces a
coupling between different degrees of freedom in equations of motions which are
highly nonlinear. In the axisymmetric halo case these systems generically have no
global integrals of motion other than energy  (in the rotating frame) and,
being far from linear, are, therefore, candidates  for supporting large (phase
space) regions of chaotic trajectories. When  a nonrotating triaxial halo is
added even energy is lost as an integral of  motion. Since time translation
symmetry is broken, such a system  has no obvious symmetries and is
likely to exhibit  chaotic  behavior from most initial conditions, 
as indeed is found to be the
case. Moreover, in  the case of a halo with a  significant core, and for mild
triaxiality, the non-axisymmetric halo perturbation is completely negligible
in the bar region.  
This is another reason why central concentration
triggers chaos; for centrally-concentrated halos of mild triaxiality the bar
and halo azimuthal forces in the equatorial plane
become comparable. There are no frames of reference where the system 
may be considered, to some approximation, stationary.

It is significant that very small deviations from axisymmetry in the halo
potential can produce drastic changes in phase space structure. 
 Systems where neither KAM stability
(a property of separable systems: see, e.g., Arnold in Mackay \& Meiss 1987) 
guarantees that small 
perturbations do not drastically change phase space structure or  
where structural stability  
(a property of strongly chaotic systems, e.g., C
systems: Anosov 1969) ensures the robustness of qualitative properties 
of trajectories, are candidates for such behavior.
In other words,  systems with 
a mixed phase space, where regular and chaotic
trajectories coexist,  are always liable for
strong modification
of qualitative properties by means of small perturbations.
This appears to be the case for models of centrally-concentrated barred
galaxies with axisymmetric halos. 
A small perturbation in the halo axis
ratio can, as we have seen,  have a significant effect (cf., Fig~\ref{moreax}). 
These systems are said to be 
structurally unstable. Modeling barred galaxies with centrally concentrated
halos while 
assuming that these halos are axisymmetric may,
therefore, produce non-generic results.

Even though we find some correlation between the energy change of               
trajectories of triaxial systems and the values of Liapunov exponents, many
trajectories that are unstable over a dynamical time conserve energy quite well
(Section~\ref{endif}).  This was inferred by calculating  the energy
correlation over times up to $5$~Gyr and the dispersion in its values averaged
over  $50$~Gyr. We find that only for a small fraction of the chaotic
trajectories does energy decorrelate over this period and the dispersion in its
values becomes large. This shows that long timescales for energy relaxation do
not imply that evolution of other quantities does not happen  at a much larger
rate, as is the case with the $z$ instability, for example.

Halos identified in 
cosmological simulations with CDM initial conditions
are found to be both
centrally-concentrated and triaxial. The results presented in this paper,
therefore, suggest that the existence of such structures around galaxies
containing strong large-scale bars is probably ruled out.  Furthermore, the
very small departures required from axisymmetry suggest that our results are
generic --- in the sense that  departures from perfect axisymmetry,
due to halo substructure, for example (another prediction of CDM models), may
be sufficient for the effects found in this work to become important.  These
findings can, therefore, be considered consistent with  the growing
observational evidence against such halos in present day galaxies.

It has been suggested  that primordial bars 
in centrally-concentrated halos can act as to reduce
the central concentration of the halo (e.g., Binney, 
Gerhard \& Silk 2000; Weinberg and Katz 2001). If anything, 
the large fraction of chaotic trajectories found here and the 
torques resulting from the interaction of two non-axisymmetric structures 
would enhance such coupling --- and that would include the effect of bar
breaking investigated by Debattista \& Sellwood (2000).

This,  however,  raises an important
question as to whether bars would form in the first place, or 
survive long enough, in unmodified CDM halos to produce
these effects. Athanassoula \& Misiriotis (2002) found that not only 
bars can form in centrally-concentrated, but axisymmetric,
halos, but that they are actually much more pronounced
that ones in less concentrated halos. The results presented 
here suggest that small non-axisymmetric perturbations
may act as to destroy these bars. Indeed, we find strong 
vertical instabilities at the outer parts of our axisymmetric 
model with centrally-concentrated halo (Model~7) which could
explain the ``X'' structure of modeled bars when seen edge
on. These vertical instabilities move inwards and become more 
pronounced when a non-axisymmetric perturbation is present.

The vertical orbital instability observed was found to proceed in tandem 
with the orbital dissolution in the $xy$-plane. If the latter is
rapid, the vertical instability will saturate and is not likely to play an
important role --- since it  almost invariably takes longer time to develop
than the instability in the plane. This is especially true if 
the initial distribution is very thin, in which case the vertical
instability timescale can be very long compared to the one in the 
plane. This is probably the reason why
Ideta \& Hozumi (2000) concluded that the vertical instability
of an $N$-body bar immersed in a centrally-concentrated halo 
does not  make  an essential contribution to its evolution.
We find that, in this case, the trajectories would
quickly mix in the $xy$-plane, evolving towards a spatial distribution which is
isotropic there and forming a lens-type configuration. 
On
the other hand, if there is sufficient time for the  vertical instability of
trajectories to develop, it can play a central role in the evolution. 
For it is especially in the intermediate cases where the instability in the plane 
may not be 
sufficient in dissolving the bar over short timescales that the additional
vertical instability may be essential (Section~\ref{shortev}).
 The final shape of the dissolved object may then be closer
in appearance to a spheroidal component.

If collective instability, leading to bar formation, is active
in the case of centrally-concentrated triaxial halos, a bar-like structure 
may appear. But if the evolution time of the density distribution
away from the barred configuration is simply
related to the Liapunov timescale of the chaotic
trajectories filling the phase space, such a  configuration will have to be
washed out in a few dynamical 
times. This should almost certainly be the situation
 in systems where the non-axisymmetric non-rotating  
 perturbation reaches $10 \%$. In this case 
there are no regular trajectories supporting the bar. Most initial
conditions lie in what appears to be 
a  connected chaotic region of phase space composed of unstable
trajectories with e-folding time of the order of $10^{8}$ Myr or smaller
(cf., Fig~\ref{histos}, bottom panel). It is also probably the case
for systems with a perturbation of $ 5 \%$ (Fig.~\ref{histos}, 
second panel from bottom). Can the presence   
of a centrally-concentrated
triaxial halos in the initial stages of galaxy evolution be the 
cause of apparent decrease in the bar fraction
at redshifts greater than 0.5 (e.g., Abraham et al. 1999)?

In the case of the ``mixed'' phase space of systems with with smaller
non-rotating perturbations the situation is more ambiguous.
Nevertheless, even a system where the non-rotating component
has a potential axis ratio as large as $0.99$ 
displays large  connected regions of initial conditions leading to strongly
chaotic trajectories.  A  comparison between the bottom panel 
of Fig.~\ref{selfcons} and middle panel of Fig.~\ref{moreax} reveals that
the configuration space  structure and orbital stability properties of such a
model is significantly different from that of the model successfully built in a
self-consistent  manner by Pfenniger (1984a,b) --- with the latter containing
far more  regular trajectories occupying a small configuration space volume
aligned  with the bar. In particular, while both models contain, at most
radii, orbits that are aligned with the bar, the variety of available such
trajectories is far larger in Pfenniger's model. Indeed, simple inspection of
the spatial structure of a sample of bar aligned orbits of Model~6 showed that
a large  fraction of these are too round to contribute towards a self 
consistent model.

The preponderance of chaotic trajectories occupying large
configuration space volumes, and the apparent absence 
of a sufficient population of regular bar-supporting orbits,
suggest that  even models
with slowly rotating non-axisymmetric perturbations of $\sim 1 \%$
cannot have time-independent equilibria.
The question arises as to 
whether the resulting time dependence would take place on 
a physically relevant timescale. This is not a trivial 
question, since in systems with such mixed phase space diffusion 
times of chaotic trajectories can sometimes be very long.
However, in the cases studied here, unless the isolated regions
of regular orbits aligned with the bar constitute a sufficient 
contribution towards a self consistent model,
the bar would have to dissolve in a timescale
significantly smaller  than a Hubble time. 
Since we know from  Section~\ref{shortev} that for  mildly triaxial 
centrally-concentrated halos the configuration space volume of most
trajectories is  also far larger than the bar's over this period.

It is possible, on the other hand, that central cusps in dark halos have been
destroyed during the  formation of the  baryonic component 
{\em via} processes  involving dynamical friction in inhomogeneous
baryon background (El-Zant, Shlosman  \& Hoffman 2001; El-Zant et al. 2002)
--- or that some fundamental physical process relating to the nature of the dark 
matter prevents the formation of central density cusps.

\acknowledgments

AEZ gratefully acknowledges hospitality of CASA and especially of Mike 
Shull. This work was supported in part by NASA grants to IS: NAG 5-10823,
WKU-522762-98-6 and HST GO-08123.01-97A.

\appendix 

\section{Evaluation of the Liapunov exponents}

\subsection{Obtaining the maximal exponent}

Even though Liapunov exponents are a widely known classical tool in nonlinear
dynamics, only  a few studies have used them to quantify chaotic
behavior in realistic galactic systems (e.g., Udry \& Pfenniger 1998;
Merritt \& Fridman 1996; Siopis \& Kandrup 2000). There is also the recent
study by Fux (2001) on the stability of trajectories in a detailed model of
the Galaxy. None give a self-contained description of how they are obtained.
We, therefore, provide such a description here, since these tools are central
to the results in this paper and in the hope that they would find a wider use
in the field.

Let a system be described by the first order 
equations of motion (Newtonian second order
equations can be replaced by two first order ones)
\begin{equation}
{\bf \dot{X}=F(\bf X},t)       \label{eq:linf1}
\end{equation}
and their variational (linearised) counterparts
\begin{equation}
{\bf \dot{\mbox{\boldmath$\xi$}}=\delta F} \label{eq:linf2}.
\end{equation}
Along a particular trajectory ${\bf \bar{X}=\bar{X}}(t,t_{o},{\bf 
\bar{X}_{o}})$ against which we would
like to measure the deviation, with ${\bf X_{o}}$ the initial conditions, 
(\ref{eq:linf2}) can be rewritten as
\begin{equation}
\dot{\mbox{\boldmath$\xi$}}={\bf D_{x}F(\bar{X}}(t,t_{o},{\bf X_{o}})),
    \mbox{\boldmath$\xi$}  
\label{eq:linf3}
\end{equation}
where ${\bf D_{x}F}$ is the Jacobian $6N \times 6N$  matrix \hspace{0.05in}
${\bf \partial F_{i}/\partial x_{j} }$ \hspace{0.05in} and  $i,j=1,6N$. 
Now let
\begin{equation}
{\bf X_{s}=X_{s}(\bar{X}}(t,t_{o},{\bf X_{o}}))
\end{equation}
be the fundamental solution of this matrix with the initial condition being
the identity matrix. The solution of~(\ref{eq:linf3}) is then given by
(Wiggins 1991) 
\begin{equation}
      \mbox{\boldmath$\xi$}={\bf X_{s}}(t)\mbox{\boldmath$\xi_{0}$},
\end{equation}
which describes the evolution under the linearised dynamics with initial
conditions $\mbox{\boldmath$\xi_{0}$}$ in the space of linear variations.

A Liapunov exponent is the infinite limit of the ``time-dependent Liapunov
exponent'' (Wiggins 1991)  at ${\bf X_{0} }$ in the direction 
${\mbox{\boldmath$\xi_{0}$}}$ at time $t$ which is given by
\begin{equation} 
\lambda(\zi,t)=\frac{\parallel
\zi(t)\parallel}{\parallel \zi_{0}
\parallel}= \frac{1}{t} \log \Bl \frac{\parallel{\bf X}_{s} (t)\zi
\parallel}{\parallel \zi_{0} \parallel}\Br \label{eq:linf4}.
\label{epn}
\end{equation}
The Liapunov exponents are then defined as
\begin{equation}
\sigma(\zi_{0}, {\bf X}_{0})=
\lim_{t \rightarrow \infty} \lambda(\zi_{0},{\bf X}_{0},t) 
\label{eq:linf5}.
\end{equation}
Numerically of course only $\lambda$ can be calculated. We will refer to the
inverse of this time-dependent Liapunov exponent as the ``exponentiation
time,'' the ``exponential timescale'' or the e-folding time.

For a Hamiltonian system with $f$ degrees of freedom, there are $2f$ 
linearly-independent directions in phase space for the vector
$\mbox{\boldmath$\xi_{0}$}$ to point at, hence there are $2f$ Liapunov
exponents. A positive Liapunov exponent indicates unstable behavior
characteristic of chaotic motion. Thus, determining the maximal exponent is
sufficient for detecting the presence of such behaviour. The evaluation of
the maximal exponent is straightforward enough. This is because exponential
instability, if it is present, will cause almost all initial linear tangent
space vectors to realign themselves along the subspace of maximal expansion. A
numerical determination of a Liapunov exponent from almost {\em any} initial
chosen direction for the linear variations  will thus tend to give an
evaluation of the maximal exponent (Wolf et al. 1985).   The only complication
that arises is that, when the exponentially increasing solutions of the
linearised equations become too large, the calculation is slowed down
(eventually leading to a numerical  overflow). This is easily remedied,
however, by application of the ``standard algorithm'' of Benettin et al.
(1976). This algorithm is based on the local averaging of the deviation
between neighboring states, which is done by dividing the time we run the
system into $n$ subintervals. An initial linearised deviation $\mbox{\boldmath
$\xi_{0}$}$ will, therefore, be transformed into 
\begin{math}
\mbox{\boldmath
$X_{s}^{1}\xi_{o},X_{s}^{2}X_{s}^{1}\xi_{o},...,X_{s}^{n}...X_{s}^{2}
X_{s}^{1}\xi_{o}$} 
\end{math}
at times $t_{1},t_{2},...,t_{n}$. At iteration $n$, the part under the
logarithm on the right hand side of~(\ref{eq:linf4}) can then be rewritten as:
\begin{equation}
\mbox{\boldmath$\parallel X^{n}_{s}...X^{3}_{s}X^{2}_{s}
    X^{1}_{s}\xi_{0}\parallel/\parallel\xi_{0}\parallel$}. 
\end{equation}
We now successively define  
\begin{equation}
\mbox{\boldmath$ \xi_{i}=X^{i}_{s}\xi_{i-1}=\parallel \xi_{i-1}
\parallel X^{i}_{s}$}\hat{\mbox{\boldmath$\xi$}}_{i-1}
\end{equation}
with  
\begin{equation}
\hat{\mbox{\boldmath$\xi$}}_{i-1}=\mbox{\boldmath$\xi_{i-1}/\parallel 
   \xi_{i-1}\parallel$}.
\end{equation}
This means that 
\begin{equation}
\mbox{\boldmath$ \parallel X_{n}...X_{2}X_{1}
\xi_{0}\parallel=\prod_{i=1}^{n}\parallel \xi_{i}\parallel $}
\end{equation}
and, therefore, if one assumes constant intervals $\Delta t$,
\begin{equation}
\lambda_{standard}=\frac{1}{\Delta t}
\lim_{n \rightarrow \infty}\sum^{i=n}_{i=1}
\frac{\log \parallel \mbox{\boldmath$\xi_{i}$}\parallel}{n}. 
\label{eq:linf6}
\end{equation}
In practice, this procedure consists of renormalizing the linearized vector 
to unity at intervals $\Delta t$, adding the logarithm of its norm to the
pre-existing  sum and restarting the integration with this renormalized unit
vector serving  as initial condition for the variational (linearised)
equations. This avoids numerical blowup.

\subsection{Obtaining the full set}

The problem of the collapse of of the linearized vectors towards the 
direction of maximum rate of expansion can be solved by re-orthogonalizing
the vectors at intervals small enough so that linear independence of the 
set of vectors is not completely lost. This can be done by repeated application 
of the Gramm-Schmidt orthogonalization procedure. Therefore, instead of just
renormalizing  one vector, as in when calculating the maximal exponent, 
at every step of the procedure described above, one {\em re-orthonormalizes} a
basis set of linearly independent tangent space vectors
$(\zi_{i}^{1}, \zi_{i}^{2}, ..., \zi_{i}^{2f})$  to obtain a new set of 
orthonormal vectors given by
\begin{eqnarray}
\acute{\zi}_{i}^{1} &=& \frac{ \zi_{i}^{1} }
{\parallel \zi_{i}^{1} \parallel }\\ 
\acute{\zi}_{i}^{2} &= & \frac{ \zi_{i}^{2}
- \Bl \zi_{i}^{2} . \acute{\zi}_{i}^{1} \Br 
\acute{\zi}_{i}^{1} } {\parallel \zi_{i}^{2}
- (\zi_{i}^{2} . \acute{\zi}_{i}^{1}) \acute{\zi} \parallel } \\
\nonumber
.\\
\nonumber
.\\
\nonumber
.\\
\acute{\zi}_{i}^{2f} &=& \frac{   \zi_{i}^{2f} - \Bl \zi_{i}^{2f} . 
    \acute{\zi}_{i}^{2f-1} \Br 
\acute{\zi}_{i}^{2f-1} -  ... -
\Bl \zi_{i}^{2f} . \acute{\zi}_{i}^{1} \Br 
\acute{\zi}_{i}^{1} }  {\parallel  \zi_{i}^{2f}-
\Bl \zi_{0}^{2f} . \acute{\zi}_{i}^{2f-1} \Br 
\acute{\zi}_{i}^{2f-1} - ...-
\Bl \zi_{i}^{2f} . \acute{\zi}_{i}^{1} \Br 
\acute{\zi}_{i}^{1} \parallel}.
\end{eqnarray}

That is the new set of vectors is made orthonormal by simply normalizing the 
first vector, then subtracting the projection of the first vector on the
second and normalizing to get the new second vector. After that, the first two
new vectors are subtracted form the third vector of the original set which
is then normalized to obtain the new third vector, etc. The vector
$\acute{\zi_{i}^{1}}$ continues to seek out the direction of maximum expansion
(since it has the same direction as $\zi$ while $\acute{\zi}_{i}^{1}$),
and $\acute{\zi}_{i}^{2}$ span the most rapidly growing two-dimensional
subspace, and, in general, the first $k$  orthogonalized  vectors span the same
subspace as the first $k$ vectors of the original set.  Also since the new set
of vectors is orthogonal, one may determine the Liapunov exponents from the
mean rate of growth of the projection  of the new vectors on the old ones.
A FORTRAN routine that finds the Liapunov exponents using this procedure is
given by Wolf et al. (1985). 

Better numerical stability can be achieved by using a {\em modified}
Gramm-Schmidt orthogonalization procedure (e.g., Lawson \& Hanson 1974; Noble
1988). This amounts to the following: instead of subtracting the $k-1$
preceding vectors from the $k^{th}$ vector, thus making the latter orthogonal
to all of the former, one starts with the $k^{th}$ vector and makes {\em all}
the following $2f-k$ vectors orthogonal to that vector. Thus we start with the
first vector to get 
\begin{eqnarray}
\acute{\zi}_{i}^{1} &=& \frac{ \zi_{i}^{1} }{\parallel \zi_{i}^{1} \parallel}\\
\zi_{i}^{2} &=& \zi_{i}^{2}-  \Bl \zi_{i}^{2}. \acute{\zi}_{i}^{1} \Br \acute{\zi}_{i}^{1}\\
\nonumber
.\\
\nonumber
.\\
\nonumber
.\\
\zi_{i}^{2f} &=& \zi_{i}^{2f}-  \Bl \zi_{i}^{2f}. \acute{\zi}_{i}^{1} \Br \acute{\zi}_{i}^{1}.
\end{eqnarray}
This is done for every vector until we get to
\begin{eqnarray}
\acute{\zi}_{i}^{2f-1} &=& \frac{ \zi_{i}^{2f-1} }{\parallel \zi_{i}^{1} \parallel}\\
\zi_{i}^{2f} &=& \zi_{i}^{2f} -  \Bl \zi_{i}^{2f} . \acute{\zi}_{i}^{2f-1} \Br \acute{\zi}_{i}^{2f-1}.
\end{eqnarray}
Here we use only this more stable modified algorithm.

\section{The Kolmogorov entropy and statistical evolution}

A discussion of the precise conditions under which the Liapunov
exponents are good indicators of statistical behavior can be found elsewhere
(e.g., Pesin 1989; Eckmann \& Ruelle 1985). We just mention that under 
certain conditions believed to be satisfied for many physical systems
the exponents are related to the Kolmogorov-Sinai entropy by
\begin{equation}
KS=\int \sum_{i} \sigma (\zi^{i}, {\bf X}_{0}) d{\bf X}_{0}
\label{EQ}
\end{equation}
where the sum is taken over all positive exponents and the integral
is over all possible initial conditions. For a single trajectory the KS
entropy (in this case simply the sum of the  positive exponents) is a 
measure of the information loss about the initial phase space point 
${\bf X}_{0}$ as the trajectory propagates, or the ``complexity'' of the 
trajectory. It is zero for regular orbits, where the motion is 
separated into recurring oscillations in each degree of freedom.
For a set of trajectories in a connected chaotic region of phase space,
it is a measure of the rate of increase of coarse-grained phase volume
they represent, and hence increase in the statistical entropy. 
It is, therefore, possible to interpret Liapunov exponents as measuring
the rate of evolution of an initially improbable distribution 
of phase space points. Nevertheless because of the existence
of phase space barriers in systems with mixed phase spaces ---
that is ones with coexisting regular and chaotic regions ---
some trajectories may take very long to reach an invariant distribution.  
The correspondence is therefore  not always so 
straightforward. It was one of our goals in this paper to test it 
for our current systems. It turns out that, for these systems,
the exponents  are very useful diagnostics of (at least) the configuration
space structure of trajectories over timescale of the order of a 
Hubble time..

\end{document}